\documentstyle[emulateapj,epsf]{article}
\def\gram{\, \hbox{g}}
\def\erg{\, \hbox{erg}}
\def\keV{\, \hbox{keV}}

\def\sec{\, \hbox{s}}
\def\day{\, \hbox{day}}
\def\days{\, \hbox{days}}
\def\year{\,\hbox{year}}
\def\Hz{\hbox{Hz}}

\def\mJy{\, \hbox{mJy}}
\def\microJy{\, \mu \hbox{Jy}}
\def\micron{\mu m}
\def\cm{\, \hbox{cm}}
\def\km{\, \hbox{km}}

\def\Mpc{\, \hbox{Mpc}}
\def\Gpc{\, \hbox{Gpc}}

\def\tE{t_{\oplus}}
\def\tEm{t_{\oplus, m}}
\def\nucm{\nu_{co, m}}
\def\nuE{\nu_\oplus}
\def\nucooE{\nu_{\oplus,{\rm cool}}}
\def\nuabsE{\nu_{\oplus,{\rm abs}}}
\def\1nE{\nu_{\oplus,1}}
\def\2nE{\nu_{\oplus,2}}

\def\nuEm{\nu_{\oplus, m}}
\def\FEm{F_{\nu, \oplus, m}}
\def\FE{F_{\nu, \oplus}}
\def\ee{{\cal E}}
\def\eemin{{\cal E}_{\rm min}}

\def\eemax{{\cal E}_{\rm max}}

\def\hf1{\hat{f}_1}
\def\fh2{\hat{f}_2}
\def\bp{Paczy\'{n}ski}
\def\mesz{M\'{e}sz\'{a}ros}
\def\iauc{IAU Circular}

\def\brsub{b}
\def\finit{f_{\brsub}}
\def\ri{r_{\brsub}}
\def\Gami{\Gamma_{\brsub}}
\def\ti{t_{\brsub}}
\def\tcoi{t_{co,\brsub}}
\def\tEi{t_{\oplus,\brsub}}
\def\rg{r_{_\Gamma}}
\def\tEf{t_{\oplus,f}}

\def\xib{\xi_{_B}}
\def\xiB{\xi_{_B}}
\def\xie{\xi_e}
\def\gemin{\gamma_{e,\hbox{\scriptsize peak}}}
\def\Fnmxe{F_{\nu,m,\oplus}}
\def\sigmaT{\sigma_{_T}}

\def\Eico{E_{i,co}}
\def\hattE{\widehat{\tE}}
\def\hatFnmxe{\widehat{\Fnmxe}}
\def\hatnuEm{\widehat{\nuEm}}
\def\Fnmxei{F_{\nu,m,\oplus,\brsub}}
\def\nuEmi{\nu_{\oplus,m,\brsub}}
\def\Eint{E_{int}}
\def\phip{\phi_p}
\def\xp{x_p}
\def\epsF{\epsilon_{_F}}
\def\epsnu{\epsilon_\nu}

\begin{document}
\title{The Dynamics and Light Curves of Beamed Gamma Ray Burst Afterglows}
\author{James E. Rhoads}
\affil{Kitt Peak National Observatory, 950 North Cherry Avenue,
Tucson, AZ 85719\altaffilmark{1}}\altaffiltext{1}{Postal address: 
 P.O. Box 26732, Tucson, AZ 85726-6732}
\begin{center}
Electronic mail: jrhoads@noao.edu \\
\smallskip
Submitted to {\it The Astrophysical Journal\/}, 2/1998;
 current revision 3/1999
\end{center}

\begin{abstract}
The energy requirements of gamma ray bursts have in past been poorly
constrained because of three major uncertainties: The distances to
bursts, the degree of burst beaming, and the efficiency of gamma ray
production.  The first of these has been resolved, with both indirect
evidence (the distribution of bursts in flux and position) and direct
evidence (redshifted absorption features in the afterglow spectrum of
GRB 970508) pointing to cosmological distances.  We now wish to
address the second uncertainty.  Afterglows allow a statistical test
of beaming, described in an earlier paper.  In this paper, we modify a
standard fireball afterglow model to explore the effects of beaming on
burst remnant dynamics and afterglow emission.  If the burst ejecta
are beamed into angle $\zeta_m$, the burst remnant's evolution changes
qualitatively once its bulk Lorentz factor $\Gamma \la 1/\zeta_m$:
Before this, $\Gamma$ declines as a power law of radius, while
afterwards, it declines exponentially.  This change results in a
broken power law light curve whose late-time decay is faster than
expected for a purely spherical geometry.  These predictions disagree
with afterglow observations of GRB 970508.  We explored several variations
on our model, but none seems able to change this result.  We therefore
suggest that this burst is unlikely to have been highly beamed, and
that its energy requirements were near those of isotropic models.
More recent afterglows may offer the first practical applications
for our beamed models.
\end{abstract}

\section{Introduction}
Understanding the energy requirements and event rates of
gamma ray bursts is necessary for any quantitative evaluation of a
candidate burst progenitor.  We need to know both how many progenitors
we expect, and how much energy they need to produce in a single event.
Until recently, both quantities were uncertain to $\sim 10$ orders of
magnitude because of the unknown distance to the bursts.  The
afterglow of GRB 970508 effectively ended that debate, because it
showed absorption lines at a cosmological redshift ($z=0.835$; Metzger
et al 1997).  This builds on earlier results from the Burst and
Transient Source Experiment (BATSE), which showed that the burst
distribution on the sky is exquisitely isotropic while the
distribution in flux is inhomogeneous (Meegan et al 1996).
These observations are best explained if the bursts are at
cosmological distances.  A very extended Galactic halo distribution
might also work, but it would have
to be unlike any other known population of Galactic objects.
The isotropy is perhaps most important now for showing that
multiple-population scenarios for gamma ray bursts cannot put any
substantial fraction of the bursters at Galactic distances.  It thus
connects the GRB 970508 redshift bound to the vast majority of the burst
population.

The dominant remaining uncertainty in the bursters' energy
requirements is now whether the bursts radiate isotropically or are
beamed into a very small solid angle.  Such beaming is allowed (though
not required) by the gamma ray observations, because the ejecta from
gamma ray bursts must be highly relativistic to explain the spectral
properties of the emergent radiation (\bp\ 1986, Goodman 1986), with
inferred minimum Lorentz factors $\Gamma \ga 100$ (Woods \& Loeb
1995).  The gamma rays we observe are therefore only those from
material moving within angle $1/\Gamma$ of the line of sight, and
offer no straightforward way of determining whether there are eject
outside this narrow cone.

These large Lorentz factors lead naturally to predictions of afterglow
emission at longer wavelengths as the burst ejecta decelerate and
interact with the surrounding material (\bp\ \& Rhoads 1993; Katz 1994;
\mesz\ \& Rees 1997a).  The characteristic frequency for this
afterglow emission depends on the Lorentz factor of the burst remnant,
and both decrease as the remnant evolves.  Such models scored a recent
triumph with the detection of X-ray, optical, and radio afterglows
from gamma ray bursts (GRBs) early in 1997 (e.g., Costa et al 1997; van
Paradijs et al 1997; Bond 1997; Frail et al 1997).  The observed
properties of the transients are in good overall agreement with the
predictions of afterglow models (Wijers, Rees, \& \mesz\ 1997; Waxman
1997a,b), although some worries remain (Dar 1997).

Because beaming depends on the relativistic nature of the flow,
afterglows can be used to test the burst beaming hypothesis.  At least
two such tests are possible.  First, because $\Gamma$ is lower at the
time of afterglow emission than during the GRB itself, the afterglow
cannot be as collimated as the GRB can.  This implies that the
afterglow event rate should exceed the GRB event rate substantially if
bursts are strongly beamed.  Allowing for finite detection thresholds,
\begin{equation} 
{N_{12} \over N_2 } \le {\Omega_1 \over \Omega_2 } \le {N_1 \over
N_{12}} ~~,
\label{ineq}
\end{equation}
where $N_1$, $N_2$ are the measured event rates above our detection
thresholds at our two frequencies; $N_{12}$ is the rate of events
above threshold at both frequencies; and $\Omega_1$, $\Omega_2$ are
the solid angles into which emission is beamed at the two frequencies.
A full derivation of this result and discussion of its application is
given in Rhoads (1997a). 

The second test is based on differences between the dynamical
evolution of beamed and isotropic bursts.  Burst ejecta decelerate
through their interaction with the ambient medium.  If the ejecta are
initially beamed into a cone of opening angle $\zeta_m$, the
deceleration changes qualitatively when the bulk Lorentz factor
$\Gamma$ drops to $1/\zeta_m$.  Prior to this, the working surface
(i.e. the area over which the expanding blast wave interacts with the
surrounding medium) scales as $r^2$.  At later times, the ejecta cloud
has undergone significant lateral expansion in its frame, and the
working surface increases more rapidly with $r$, eventually
approaching an exponential growth.  Spherical symmetry prevents this
transition from occurring in unbeamed bursts.  A brief analysis of
this effect was presented in Rhoads (1997b).

We have two major aims in this paper.  First, we will present a full
derivation of the late time burst remnant dynamics for a beamed gamma
ray burst.  We support this by calculating the emergent synchrotron
radiation for two electron energy distribution models, but we do not
attempt to do so for all possible fireball emission scenarios.
Second, we observe that our model is not consistent with any
small-angle beaming of GRB 970508.  This implies a substantial minimum
energy for this burst.  If radiative efficiencies are lower than $\sim
10\%$, this limit approaches the maximum energy available in compact
object merger events.  We explore possible ways to evade this minimum
energy requirement through other forms of beaming models, but find
none.  We therefore conjecture that such models cannot be constructed
for GRB 970508
unless the usual fireball model assumptions about relativistic blast
wave physics are substantially modified, and challenge the community
to prove this assertion right or wrong.

We explore the dynamical evolution of a model beamed burst in
section~\ref{dynamics}.  In section~\ref{lightcurve} we incorporate a
model for the electron energy spectrum and magnetic field strength and
so predict the emergent synchrotron radiation.  In
section~\ref{discuss}, we compare the model with observed
afterglows.  The early (1997) data appeared inconsistent with the
beaming model, suggesting that bursts are fairly isotropic and
therefore very energetic events.  Finally, in section~\ref{min_en}, we
explore variations on our model to try to reduce the inferred energy
needs of GRB 970508.  We comment briefly on more recent data and
summarize our conclusions in section~\ref{summary}.

\section{Dynamical Consequences of Beaming} \label{dynamics}
We explore the effects of beaming on burst evolution using the
notation of~\bp\ \& Rhoads (1993).  Let $\Gamma_0$ and $M_0$ be the
initial Lorentz factor and ejecta mass, and $\zeta_m$ the opening
angle into which the ejecta move.  The burst energy is $E_0 = \Gamma_0
M_0 c^2$.  Let $r$ be the radial coordinate in the burster frame; $t$,
$t_{co}$, and $\tE$ the time from the event measured in the burster
frame, comoving ejecta frame, and terrestrial observer's frame; and
$f$ the ratio of swept up mass to $M_0$.

The key assumptions in our beamed burst model are that (1) the initial energy
and mass per unit solid angle are constant at angles $\theta <
\zeta_m$ from the jet axis and zero for $\theta > \zeta_m$; (2) the
total energy in the ejecta + swept-up material
is approximately conserved; (3) the ambient
medium has uniform density $\rho$; and (4) the cloud of ejecta +
swept-up material expands in its comoving frame at the sound speed
$c_s = c/ \sqrt{3}$ appropriate for relativistic matter.  The last of
these assumptions implies that the working surface of the expanding
remnant has a transverse size $\sim \zeta_m r + c_s t_{co}$.  The
evolution of the burst changes when the second term dominates over the
first.

Each of these assumptions may be varied, but we believe the
qualitative change in burst remnant evolution will remain over a wide
range of possible beaming models.  Removing assumption~(4) is the only
obvious way to turn off the dynamical effects of beaming, and even
then observable breaks in the light curve are expected when $\Gamma
\sim 1/\zeta_m$.

There are several models in the literature that use radiative rather
than adiabatic models, dropping our second assumption.  The case for
radiative bursts depends on the efficiency with which relativistic
shocks transfer bulk kinetic energy to magnetic fields and electrons,
and I regard the validity of assumption~(2) as an open question.  For
a closer examination of this issue, I refer the reader to papers by
Vietri (1997a,b) and by Katz \& Piran (1997a), who advocate radiative
models; and to Waxman, Kulkarni, \& Frail (1998), who defend the
adiabatic model.  \mesz, Rees, \& Wijers (1997) point out that the
dynamical consequences ($\Gamma \propto r^{-3}$) of radiative models
depend on equipartition between protons, electrons, and magnetic
fields being maintained at all times.  Thus, a short electron cooling
time will affect the afterglow radiation, but will not necessarily
result in $\Gamma \propto r^{-3}$.  Sari (1997) considers corrections
to the adiabatic burst evolution for modest energy losses.

Models that do not use assumption~(1) have been discussed by \mesz,
Rees, \& Wijers (1997) and Panaitescu, \mesz, \& Rees (1998).
Finally, assumption~(3) has been dropped by several authors (Vietri
1997b; \mesz, Rees, \& Wijers 1997; Panaitescu, \mesz, \& Rees 1998)
in favor of a more general power law density $\rho \propto r^{-g}$.
Such models complicate the beamed burst analysis and will change the
form of $\Gamma(r)$ relation but will leave intact the basic
conclusion that $\Gamma(r)$ changes qualitatively when $\Gamma \la
1/\zeta_m$.

\subsection{Dynamical Calculations: Numerical Integrations}
Given these assumptions, the full equations describing the burst
remnant's evolution are
\begin{equation}
f = {1\over M_0} \int_0^r r^2 \Omega_m(r) \rho(r) dr ~~,
\end{equation}
\begin{equation} \label{omega_m_def}
\Omega_m \approx \pi ( \zeta_m + c_s t_{co} / c t )^2
 \approx \pi ( \zeta_m +  t_{co} / \sqrt{3} t )^2 ~~,
\end{equation}
\begin{equation} \label{gam_of_f}
\Gamma = \left( \Gamma_0 + f \right) / \sqrt{ 1 + 2 \Gamma_0 f + f^2}
\approx \sqrt{ \Gamma_0 / 2 f } ~~,
\end{equation}
 %
\begin{displaymath}  
 t = r/c~~, \quad 
t_{co} = \int_0^t  dt' / \Gamma ~~,
\end{displaymath}
\begin{equation} \label{timedefs}
\quad \hbox{and} \quad
\tE = (1+z) \int_0^t  dt' / 2  \Gamma^2 ~~.
\end{equation}
Equation~\ref{gam_of_f} is derived in \bp\ \& Rhoads (1993) from
conservation of energy and momentum, along with algebraic
simplifications of equations~\ref{timedefs} for the spherical case.
The definition of $\tE$ here includes the cosmological time
dilation factor $(1+z)$ for a source at redshift $z$.
Equation~\ref{omega_m_def} is not strictly valid when $\zeta_m \ga 1$,
but we will accept this deficiency since the error thereby introduced
is not a dominant uncertainty in our results.

These equations can be solved by numerical integration to yield
$f(r)$, $\Gamma(r)$, and $\tE(r)$.  Figure~1 shows $\Gamma(r)$ from such
integrations for an illustrative pair of models (one beamed, one
isotropic).

\begin{figure*}[h!] 
\epsfxsize=\hsize \epsfbox{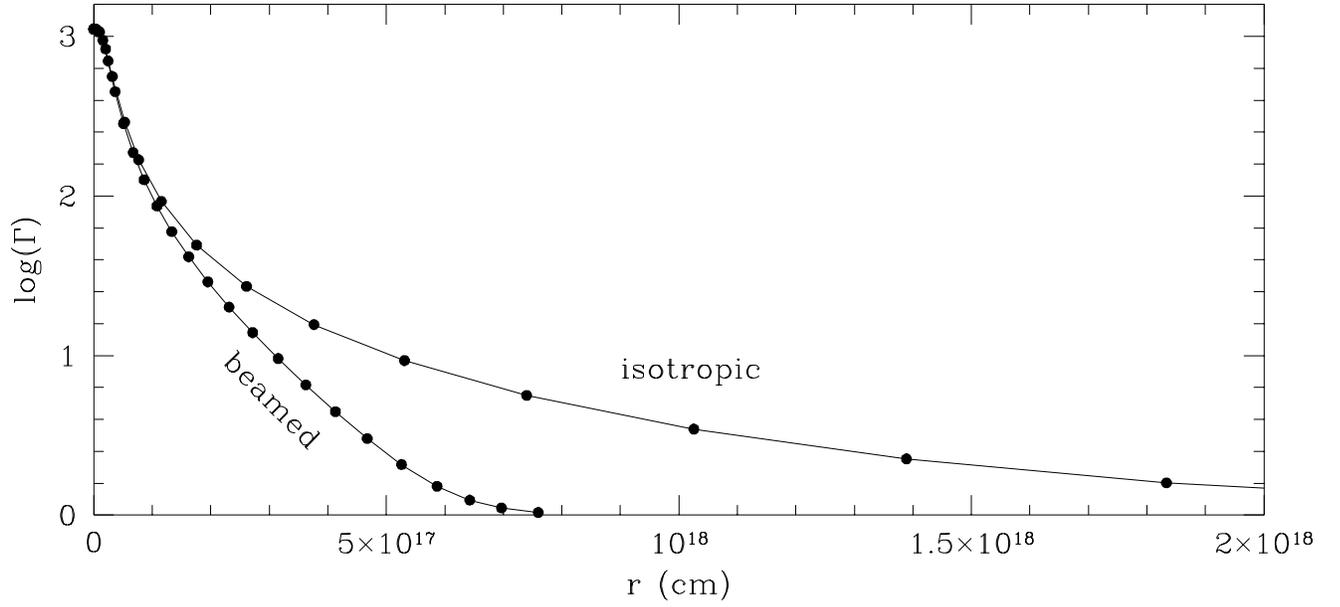}
\vspace{10pt}
\caption{Dependence of the bulk Lorentz factor $\Gamma$ on the burst
expansion radius for an isotropic burst and a burst beamed into an
opening angle $\zeta_m = 0.01$ radian.  Both bursts follow a $\Gamma
\propto r^{-3/2}$ evolution initially, but the beamed burst changes its
behavior at
$\Gamma \approx 100 \approx 1/\zeta_m$, beyond which its Lorentz factor
decays exponentially with radius.
}
\label{gamrfig}
\end{figure*}

\subsection{Dynamical Calculations: Analytic Integrations}
The most interesting dynamical change introduced by beaming is a
transition from a power law $\Gamma \propto r^{-3/2}$ to an
exponentially decaying regime $\Gamma \propto \exp(-r/r_{_\Gamma})$.
We will first give a derivation of the power law behavior.

\subsubsection{Power Law Regime} \label{pl_dyn_analytic}
Consider the approximate evolution equations for the regime where
(a) $1/\Gamma_0 \la f \la \Gamma_0$, so that $\Gamma \approx
\sqrt{\Gamma_0 / 2 f }$; and (b) $c_s t_{co} \la \zeta_m r$
(corresponding to $f \la 9 \Gamma_0 \zeta_m^2$).
\begin{eqnarray} \nonumber
d f / d r & \approx & {\pi \over M_0} \left( \zeta_m r \right)^2 \rho
\quad , \quad
d t_{co} / d r  \approx \sqrt{2 f \over c^2 \Gamma_0} \quad , \\
d \tE / d r & \approx & (1+z) { f \over c \Gamma_0 } \quad .
\label{eqn_powlaw1}
\end{eqnarray}
The initial conditions are $f = 0$ and $t = t_{co} = \tE = 0$ at
$r=0$.  So we can easily integrate and obtain
\begin{equation}
f = {\pi \zeta_m^2 \over 3} {\rho \over M_0} r^3  ~~,
\end{equation}
\begin{equation}
\Gamma \approx \left( 3 M_0 \Gamma_0 \over 2 \pi \zeta_m^2 r^3 \rho
\right)^{1/2} =  \left( 3 E_0 \over 2 \pi \zeta_m^2 c^2 r^3 \rho \right)^{1/2}
\end{equation}
\begin{equation}
t_{co} = \left( 8 \pi  \zeta_m^2 \rho \over 75 E_0 \right)^{1/2} r^{5/2}
 = {2 \over 5} {r \over c \Gamma}
\label{tco_r_pow}
\end{equation}
\begin{equation}
\tE 
 = (1+z) {\pi \zeta_m^2 \over 12} {c \rho \over E_0} r^4
\end{equation}
whence
\begin{equation} \label{gam_tE_pl}
\Gamma = {2^{-5/4}} \left( 3 E_0 \over \pi \zeta_m^2 c^2 \rho
\right)^{1/8} \left( 1 + z \over c \tE \right)^{3/8} ~~.
\end{equation}
By making the substitutions $\pi \zeta_m^2 \rightarrow 4 \pi$ and
$(1+z) \rightarrow 1$ in these results, we recover the evolution of a
spherically symmetric burst remnant derived by \bp\ \& Rhoads (1993).

\subsubsection{Exponential Regime} \label{exp_dyn_analytic}
To demonstrate the exponential behavior, consider the approximate
evolution equations for the regime where (a) $1/\Gamma_0 \la f \la
\Gamma_0$, so that $\Gamma \approx \sqrt{\Gamma_0 / 2 f }$; and (b)
$c_s t_{co} > \zeta_m r$ (corresponding to $f \ga 9 \Gamma_0
\zeta_m^2$):
\begin{eqnarray} \nonumber
d f / d r & \approx & {\pi \over M_0} c_s^2 t_{co}^2 \rho \quad , \quad
d t_{co} / d r  \approx \sqrt{2 f \over c^2 \Gamma_0} \quad ,\\
d \tE / d r & \approx& (1+z) { f \over c \Gamma_0 } \quad .
\label{eqn_simpler}
\end{eqnarray}
By forming the ratio $(d f / d r) / (d t_{co} / d r)$ and isolating
terms with $f$ and with $t_{co}$, it follows that 
\begin{equation}
\sqrt{f} d f = {\pi \over \sqrt{2}} { c\, c_s^2 \rho \sqrt{\Gamma_0}
\over M_0} \times  t_{co}^2 d t_{co}
 \approx  {\pi \over 3 \sqrt{2}} { c^3 \rho \sqrt{\Gamma_0}
\over M_0} \times t_{co}^2 d t_{co} ~~.
\end{equation}
This is easily integrated to obtain
\begin{equation}
f^{3/2} = \left( \pi \sqrt{\Gamma_0} c\, c_s^2 \rho \over \sqrt{8}  M_0
\right) \left( t_{co}^3  - c_1 \right) ~~,
\label{eqn_f_tco}
\end{equation}
where $c_1$ is a constant of integration.  Using the initial
conditions for the exponential regime derived below
(eqn.~\ref{initc_first} -- \ref{initc_last}), one can show that the
constant of integration is
$c_1 =  -25 E_0 \zeta_m^3 / (4 \pi \rho c_s^5 )$,
which becomes negligible once $c_s t_{co} \gg \zeta_m r$.  
Equation~\ref{eqn_f_tco} then becomes $f \propto t_{co}^2$, and 
we see  from equations~\ref{eqn_simpler} and~\ref{gam_of_f} that $f$,
$\Gamma$, $t_{co}$, and $\tE$ will all behave exponentially with $r$ in
this regime. 
Retaining the constants of proportionality, we find
\begin{eqnarray} \nonumber
f & \propto & \exp( 2 r / r_{_\Gamma} ) \quad \hbox{where} \\
\rg & = &\left[ {1 \over \pi} \left(c \over c_s \right)^2 
{ \Gamma_0 M_0 \over \rho } \right]^{1/3}
 = \left[ E_0 \over \pi c_s^2 \rho \right]^{1/3} ~~.
\end{eqnarray}
Further algebra yields $\Gamma \propto \exp(-r / r_{_\Gamma} )$,
$t_{co} \propto \sqrt{f} \propto \exp(r/\rg)$, and 
$\tE \propto f \propto \exp( 2 r /  r_{_\Gamma} )$,
so that $\Gamma \propto \tE^{-1/2}$.
Thus, while the evolution of $\Gamma(r)$ changes from a power law to an
exponential at $\Gamma \sim 1/\zeta_m$, the evolution of $\tE(r)$
changes similarly.  The net result is that $\Gamma(\tE)$ has a power
law form in both regimes, but with a break in the slope from $\Gamma
\propto \tE^{-3/8}$ when $\Gamma >  1/\zeta_m$ to $\Gamma \propto
\tE^{-1/2}$ when $\Gamma < 1/\zeta_m$.

The initial conditions for the exponential regime are approximately
set by inserting the transition condition  $c_s t_{co} = \zeta_m c t$
into the evolution equations for the power law regime. 
Denoting the values at this break with the subscript $_\brsub$, we have
$c_s \tcoi =  \zeta_m c \ti = \zeta_m \ri$, which we combine with
equation~\ref{tco_r_pow} to obtain 
\begin{equation}
\ri = \left( 75 E_0  \over 8 \pi \rho c_s^2 \right)^{1/3} \qquad
\hbox{and} \qquad \finit = {25 \over 8} \left( c \over c_s \right)^2 
\zeta_m^2 \Gamma_0 ~~.
\label{initc_first}
\end{equation}
The corresponding values for $\Gamma$, $\tE$, and $t_{co}$ are
\begin{eqnarray}
\Gami & = & {2 c_s \over 5 c } { 1 \over \zeta_m} \quad , \label{initc_Gam}\\
\tEi & = & (1+z) \left( 3 \over \pi \right)^{1/3} { 5^{8/3} \over 64 } {c \over c_s}
\left( E_0  \over \rho c_s^5 \right)^{1/3} \zeta_m^2 ~~ ,~
\label{initc_tE} \\
\hbox{and} \quad
 \tcoi & = & \zeta_m \left( 75 E_0 \over 8 \pi \rho c_s^5 \right)^{1/3}
\label{initc_last}
\end{eqnarray}

The evolution in the exponential regime is then approximated by
\begin{equation}
\begin{array}{rcrcl}
f & = & \finit & \times & \exp\left\{ 2 (r-\ri)/\rg\right\} \\
\tE & = & \tEi & \times & \exp\left\{ 2 (r-\ri)/\rg\right\} \\
\Gamma & = & \Gami & \times & \exp\left\{-(r-\ri)/\rg\right\} \\
t_{co} & = & \tcoi & \times & \exp\left\{ (r-\ri)/\rg\right\}
\end{array}
\label{expdyn}
\end{equation}

A thought experiment that will help understand the onset of the
exponential decay of $\Gamma$ with radius is to consider the shape of
a GRB remnant in a pressureless, uniform ambient medium at late times
(after all motions have become nonrelativistic).  In the spherical
case, the blast wave will leave behind a spherical cavity.  In the
beamed geometry, the cavity will be conical near the burster, but will
change shape at the radius where the lateral expansion of the remnant
becomes important.  At this point, the cone flares, and the mass swept
up per unit distance begins to grow faster than $r^2$.  This
corresponds to the onset of the exponential $\Gamma(r)$ regime.  The
cone continues to become rapidly wider until it reaches the radius
where the remnant becomes nonrelativistic. The final cavity resembles
the bell of a trumpet.  It is unclear whether such remnants would
survive long enough to be observed in a realistic interstellar medium.

\section{Emergent Radiation} \label{lightcurve}
The Lorentz factor $\Gamma$ is not directly observable, and we
ultimately want to predict observables like the frequency of peak
emission $\nuEm$, the flux density $\FEm$ at $\nuEm$, and the angular
size $\theta$ of the afterglow.  To do so, we need to introduce a
model for the emission mechanism.  We will restrict our attention to
synchrotron emission, which is the leading candidate for GRB afterglow
emission.  We first consider the case of optically thin emission with
a steep electron energy spectrum.  This emission model is used
in many recent afterglow models (e.g. Waxman 1997a,b).  We then repeat the
calculation for the emission model of \bp\ \& Rhoads (1993).

\subsection{General equations: Optically thin case}
Our dynamical model for burst remnant evolution gives the volume $V$
and internal energy density $u_i$ of the ejecta as a function of
expansion radius $r$.  Detailed predictions of synchrotron emission
require the magnetic field strength and the electron energy spectrum.
We assume that the energy density in magnetic fields and in
relativistic electrons are fixed fractions $\xib$ and $\xie$ of the
total internal energy density.  The magnetic field strength $B$
follows immediately:
\begin{equation}
B = \sqrt{ 8 \pi \xib u_i} ~.
\end{equation}
(N.b., we use the notation of
\bp\ \& Rhoads 1993.  Some other authors have instead defined $\xib$ in 
terms of the magnetic field strength, such that $B \propto \xib$ in their
models; care must therefore be taken in comparing scaling laws under
these alternative notations.)

The electron energy spectrum requires additional assumptions.  We
first follow Waxman's (1997a,b) assumptions, to facilitate comparison
of our results for beamed bursts with his for unbeamed bursts.  In the
frame of the expanding blast wave, the swept-up ambient medium
appears as a relativistic wind having Lorentz factor $\Gamma$.  We
assume that the electrons from the ambient medium have their direction
of motion randomized in the blast wave frame.  Moreover, they may
achieve some degree of equipartition with the protons.  The typical
random motion Lorentz factor $\gamma_e$ for the swept-up electrons in
the blast wave frame is then in the range $\Gamma \la \gamma_e \la 0.5
(m_p/m_e) \Gamma$.  In terms of the energy density fraction in
electrons, $\langle \gamma_e \rangle \approx \xie (m_p/m_e) \Gamma$.
We further assume that the electrons in the original ejecta mass are
not heated appreciably, so that the number of relativistic electrons
is $N_e = f M_0 / (\mu_e m_p)$ (where $\mu_e$ is the mean molecular
weight per electron) rather than $(1+f) M_0 / (\mu_e m_p)$.  We take
the electron energy $\ee$ to be distributed as a power law $N(\ee) =
\ee^{-p}$ for $\eemin < \ee < \eemax$, where $N(\ee) d\ee$ is the
number of electrons with energies between $\ee$ and $\ee + d\ee$.
Finally, we assume that $p>2$, so that the total electron energy
$\int_{\eemin}^{\eemax} \ee N(\ee) d\ee$ is dominated by electrons
with $\ee \approx \eemin$, and $\gemin \approx \langle \gamma_e
\rangle \approx \xie (m_p/m_e) \Gamma \approx \eemin / (m_e c^2)$.

The optical depth to synchrotron self-absorption is assumed to be
small at the characteristic synchrotron frequency corresponding to
$\eemin = \gemin m_e c^2$.  In the comoving frame, this frequency is
$\nucm = 0.29 \times 3/(4\pi) \langle \sin\alpha \rangle \gemin^2 e B / ( m_e c )
= 0.29 \times (3/16)  \gemin^2 e B / ( m_e c )$ (Pacholczyk 1970, 
Rybicki \& Lightman 1979), where the calculation of the mean pitch
angle $\langle \sin\alpha \rangle = \pi/4$ assumes an isotropic distribution
of electron velocities and a tangled magnetic field.  Wijers \& Galama (1998)
have integrated over the power law distribution of electron energies to show
that the peak comoving frame frequency for a power law energy distribution
becomes $\langle \nucm \rangle_\ee 
=3 \xp / (4\pi) \times \gemin^2 e B / ( m_e c )$, where $\xp$ is
a function of the power law index $p$, and where $0.64 \ga \xp \ga 0.45$ 
for $2 < p < 3$.  Below this peak frequency, the flux
density rises as $\nu^{1/3}$, while at higher frequencies it falls as
$\nu^{-\alpha}$ where $\alpha = (p-1)/2$ (e.g., Rybicki \& Lightman
1979, Pacholczyk 1970).  Three additional breaks may occur,
corresponding to the highest electron energy attained in the shock, 
the electron energy above which cooling is important, and the frequency
where synchrotron self-absorption becomes important.  We will comment
on the cooling break below, and will ignore the other two breaks
for the present.

We first estimate the observer-frame frequency $\nuEm$ at which the
spectrum peaks.  
This is
\begin{eqnarray} \nonumber
\nuEm & \approx & \langle 1 + \beta \cos(\theta)\rangle_\theta
 \Gamma \times \langle \nucm \rangle_\ee /  (1+z) \\
\label{nu_max}
& \approx & {1\over 1+z} \times {4 \Gamma \over 3} \times
 { 3 \xp \over 4\pi } \times {\gemin^2 e B \over  m_e c } \\
\nonumber & \approx & {1\over 1+z} {\xp \over \pi}
\left( \xie {m_p \over m_e} \right)^2 {e B \over m_e c }  \Gamma^3
\end{eqnarray}
where the factor $\langle 1 + \beta \cos(\theta) \rangle \Gamma$
is the Lorentz transformation for frequency,
and where $\beta = \sqrt{1-\Gamma^{-2}}$ is the
expansion velocity as a fraction of lightspeed $c$.  $\theta$ is the
angle between the velocity vector of radiating material and
the photon emitted, as measured in the frame of the emitting matter.
$\langle 1 + \beta\cos(\theta)\rangle$ denotes an average weighted
by the received intensity.  We shall use the highly relativistic limit
$\beta \rightarrow 1$ throughout this work, which leads to the result 
$\langle 1 + \beta\cos(\theta)\rangle = 4/3$ applied in the final
lines of equation~\ref{nu_max} (Wijers \& Galama 1998).

We estimate the peak flux density following equations~19 and 25
of Wijers \& Galama (1998). 
The basic equation is
\begin{equation}
\Fnmxe = \Gamma \times N_e \times \phip {\sqrt{3} e^3 B \over m_e c^2}
\times {1+z  \over \Omega_\gamma d^2 }~~.
\label{fnu_basic1}
\end{equation}
Here $\phip {\sqrt{3} e^3 B/ (m_e c^2)}$ is the average comoving frame
peak luminosity per unit frequency emitted by a single electron.  The
details of the average over pitch angle and electron energy are hidden
in the factor $\phip$, which is a function of the electron energy
distribution index $p$ with range $0.59 \la \phip \la 0.66$ for $2 < p
< 3$ (Wijers \& Galama 1998).  The factor $\Gamma$ accounts for the
Lorentz transformation of flux density (Wijers \& Galama 1998).
Distance and beaming effects enter through the factor
$1/(\Omega_\gamma d^2)$, where $d$ is the luminosity distance to the
burst, and $ \Omega_\gamma$ is the solid angle into which radiation is
beamed.  Finally, redshift affects the flux density by factor $(1+z)$
(e.g. Weedman 1986).
Our goal is to express
$\Fnmxe$ purely in terms of the dynamical variables we calculated in
section~\ref{dynamics}.

The comoving frame synchrotron cooling time is
\begin{equation}
t_s = { 6 \pi m_e c \over \sigmaT \gamma_e B^2} \approx
 { 6 \pi m_e c \over \sigmaT \gemin B^2}~~,
\end{equation}
where $\sigmaT = 0.665\times 10^{-24} \cm^2$ is the Thompson
cross-section (e.g., Rybicki \& Lightman 1979).  

To obtain the
magnetic field strength $B$, we need the volume of the ejecta cloud,
which will have transverse radius $\sim r \zeta_m + c_s t_{co}$ and
thickness $\sim c_s t_{co}$, giving comoving volume $V=\pi (c t \zeta_m +
c_s t_{co})^2 (c_s t_{co})$.  Under the approximation of negligible
radiative losses, the internal energy is given by $\Eico = 
E_0 / \Gamma$.  The comoving frame magnetic field strength is thus
\begin{equation}
B = \left( 8 \xib E_0 \over \Gamma (c t \zeta_m + c_s t_{co})^2 (c_s
 t_{co}) \right)^{1/2} ~~.
 \label{B_co}
\end{equation}

The remaining pieces are trivial.
$\Omega_\gamma \approx \pi \left(\zeta_m + 1/\Gamma \right)^2$,
and $d$ and $1+z$ are simply scale factors.

We now have all the pieces of equations~\ref{nu_max}
and~\ref{fnu_basic1} expressed in terms of dynamical variables from
section~\ref{dynamics}.  This means that we can insert these formulae
into our numerical integration code and calculate $\Fnmxe$ and $\nuEm$
as a function of $\tE$ (or of $f$, $\Gamma$, or $r$).
In order to determine a light curve at fixed observed frequency, we
combine the broken power law spectral shape described above with the
calculated frequency and flux density of the spectral peak to
determine the approximate flux density at the observed frequency and
time.

The cooling break and self absorption break (cf. Sari, Piran, \&
Narayan 1998) are additional observed features in afterglow data.  We
do not treat either in detail here, but do we present a derivation of
the cooling break behavior for beamed gamma ray bursts elsewhere
(Rhoads 1999b).  These results are summarized below.  We have not yet
treated the self-absorption break.  Self-absorption is important
primarily at low frequencies, where scintillation can hamper light
curve slope measurements.  For a treatment of this regime in beamed
bursts, see Sari, Piran, \& Halpern (1999).

Finally, we consider the evolution of the apparent angular size
$\theta$.  In the spherical case or the power-law regime for a beamed
burst, $\theta = r/(\Gamma d_\theta) \propto \tE^{5/8}$ (where $
d_\theta$ is the angular diameter distance to the burst).  In the
exponential regime, $\theta$ is determined by the physical transverse
size of the ejecta cloud rather than the beaming angle, but the
difference is not dramatic because the physical size increases at $c_s
\sim c$.  The result is therefore $\theta \approx c_s t_{co} /
d_\theta \propto 1/\Gamma \propto \tE^{1/2}$.  There is also an
intermediate regime, valid for the brief time when $1/\Gamma \ga
\zeta_m \ga (c_s t_{co}) / (c t )$.  In this case,  $\theta \propto
\zeta_m r \propto \tE^{1/4}$.  If the exponential regime did not
happen at all, this behavior would continue for all $\Gamma <
1/\zeta_m$.

\subsection{Analytic Results: Optically thin case} \label{analytic_othin}
In the limiting cases where one of the terms in the transverse size
$\zeta_m c t + c_s t_{co}$ is dominant and the other negligible, we
can derive analytic expressions for $\Fnmxe$ and $\nuEm$ as functions
of observed time  $\tE$ and the physical parameters of the fireball.
We begin with the early time case, and show that its light curve is
observationally indistinguishable from that of an isotropic burst.

\subsubsection{Power Law Regime}
We first determine the comoving magnetic field in this regime
by inserting $\zeta_m c t \gg c_s t_{co}$ into equation~\ref{B_co}
to obtain 
\begin{equation} \label{Bcopow}
B = {2^{1/4} \sqrt{5 \pi} \over 3^{3/8} } {c^{7/8} \over c_s^{1/2}}
\xiB^{1/2}\left(E_0 \over \pi \zeta_m^2 \right)^{1/8} \left( \rho (1+z)
\over \tE \right)^{3/8} ~~.
\end{equation}
Inserting this result into equation~\ref{nu_max}, we find
\begin{eqnarray}
\nonumber \nuEm & = & {2^{1/4} 5^{1/2} \over 3^{3/8} \pi^{1/2}}
 \xp \xie^2 \xib^{1/2} \left( m_p \over m_e \right)^{2}
{ e \over m_e c^{1/8} c_s^{1/2} } \\ 
 & \times & \left(E_0 \over \pi \zeta_m^2 \right)^{1/8}
\left(\rho \over \tE \right)^{3/8} \Gamma^3 (1+z)^{-5/8}\\
\nonumber &  = & {5^{1/2} \over 2^{7/2} }\xp \xie^2 \xib^{1/2}
 \left( m_p \over m_e \right)^{2} { e \over m_e c^{2} c_s^{1/2} } \\
 & \times & \left(E_0 \over \pi \zeta_m^2 \right)^{1/2} \tE^{-3/2} (1+z)^{1/2}
\label{numax_powlaw}
\end{eqnarray}
where we have used equation~\ref{gam_tE_pl} to eliminate $\Gamma$
in the last line.

For the cooling break, we obtain $\nucooE \propto \tE^{-1/2}$ (Rhoads
1999b; Sari et al 1999).

Turning our attention to $\Fnmxe$,
we first need the number $N_e$ of radiating electrons in terms of
$\tE$.  For the power law regime, this becomes
\begin{equation}
N_e = {2^{3/2} \over 3^{1/4}} { \left( \pi \zeta_m^2 \rho \right)^{1/4}
\over \mu_e m_p } \left( E_0 \tE \over c (1+z) \right)^{3/4}~~.
\end{equation}
Combining this with equations~\ref{fnu_basic1}, \ref{Bcopow}, 
and the appropriate limiting form of $\Omega_\gamma$, we find
\begin{equation}  \label{fnu_powlaw}
\Fnmxe = \sqrt{10 \pi} {\phip \xib^{1/2} \over \mu_e m_p}
{e^3 \over m_e c^3} \sqrt{ c \over c_s} {\rho^{1/2} E_0
\over \pi \zeta_m^2 } {1+z \over d^2} ~.
\end{equation}
Note that this result is independent of $\tE$.

Apart from small differences in the numerical coefficients,
our results for the power law regime are essentially
the same as the results that Waxman (1997a,b) and Wijers
and Galama (1998) obtained for isotropic bursts.  Differences between
our results and Waxman's are primarily because we have adopted the more
precise treatment of the synchrotron peak frequency presented
by Wijers and Galama (1998), while differences between our
results and those of Wijers and Galama stem from a slightly
different way of calculating the comoving frame magnetic field.

\subsubsection{Exponential Regime}
When $c_s t_{co} \gg \zeta_m c t$, we are in the regime where
$\Gamma$, $\tE$, etc.\ all behave exponentially with radius
(section~\ref{exp_dyn_analytic}).   We first
rewrite the scalings from equation~\ref{expdyn} as
\begin{equation}
\begin{array}{rclcl}
\tE / \tEi & = & \exp\left\{ 2 (r-\ri)/\rg\right\} & & \\
f / \finit & = & \exp\left\{ 2 (r-\ri)/\rg\right\} & = & \tE / \tEi \\
\Gamma / \Gami & = & \exp\left\{-(r-\ri)/\rg\right\} & = &
 ( \tE / \tEi )^{-1/2} \\
t_{co} / \tcoi & = & \exp\left\{ (r-\ri)/\rg\right\} & = &
 ( \tE / \tEi )^{+1/2} ~.
\end{array}
\label{expdyn2}
\end{equation}

We next determine the comoving frame magnetic field in the appropriate limit:
\begin{equation} \label{Bcoexp}
B = \left( 8 \xib E_0 \over c_s^3 \right)^{1/2} \Gami^{-1/2}
 \tcoi^{-3/2} \left( \tE \over \tEi \right)^{-1/2} ~~.
\end{equation}
Combining this result with equation~\ref{nu_max}
allows us to determine the peak frequency as
\begin{eqnarray}
\nonumber  \nuEm &=& {1 \over 1+z}
{2^{3/2} \over \pi} \xp \xie^2 \xib^{1/2} \left(m_p
\over m_e \right)^2 {e \over m_e c } \\
 & \times  &  E_0^{1/2}
\Gami^{5/2} (c_s \tcoi)^{-3/2} \left(\tE \over \tEi \right)^{-2} ~~.
\end{eqnarray}
Substituting for the exponential regime initial conditions
from equations~\ref{initc_first}--\ref{initc_last} then yields
\begin{eqnarray} 
\nonumber \lefteqn{  \nuEm = {1 \over 1+z}
{ 2^{11/2} \over 3^{1/2} 5^{7/2} \pi^{1/2} }
\xp \xie^2 \xib^{1/2} \left( m_p \over m_e \right)^2  }  \\
& & \mbox{} \times {e \over m_e } \left(c_s \over c \right)^{7/2}
{\rho^{1/2} \over \zeta_m^{4}} \left( \tE \over \tEi
\right)^{-2} \label{numax_exp} \\
\nonumber & = & (1+z) {3^{1/6} 5^{11/6} \over 2^{13/2} \pi^{7/6}}
\xp \xie^2 \xib^{1/2} \left(m_p \over m_e \right)^2 \\
& & \mbox{} \times {e \over m_e} \left(c_s \over c \right)^{3/2}
{E_0^{2/3} \over \rho^{1/6} c_s^{10/3} } \tE^{-2} ~~.
\end{eqnarray}

The observed cooling break frequency ceases to evolve in this regime:
$\nucooE \propto \tE^0$ (Rhoads 1999b; Sari et al 1999).

Turning now to the amplitude of the spectral peak, we
combine
\begin{equation}
N_e = {25 \over 8} {E_0 \zeta_m^2 \over
c_s^2 \mu_e m_p } {\tE \over \tEi}
\end{equation}
and
\begin{equation}
\Omega_\gamma = \pi  \Gamma^{-2} =
\pi \Gami^{-2} (\tE / \tEi)
\end{equation}
with equations~\ref{fnu_basic1} and \ref{Bcoexp}
to obtain
\begin{eqnarray}
\Fnmxe & = & {3^{1/2} 5^2 \over 2^{3/2} \pi }
\phip \xib^{1/2} { e^3 \over m_e m_p \mu_e}
{E_0^{3/2} \zeta_m^2 \over c^2 c_s^{2}} \\
& \times & (c_s \tcoi)^{-3/2} \Gami^{5/2}
\left( \tE \over \tEi \right)^{-1} 
{1+z \over d^2}~~.
\end{eqnarray}
Substituting the initial conditions for the exponential
regime, this becomes
\begin{eqnarray} \label{fnu_exp}
\Fnmxe & = & {\sqrt{32 \pi \over 125} } \phip
\xib^{1/2} { e^3 \over c^3 m_p m_e \mu_e }\left(c_s
\over c \right)^{3/2} \\
\nonumber & & \mbox{} \times{ E_0 \sqrt{\rho} \over \pi \zeta_m^2}
\left( \tE \over \tEi \right)^{-1} {1+z \over d^2} \\
& = & {3^{1/3} 5^{7/6} \over 2^{7/2} \pi^{5/6} }
\phip \xib^{1/2} {e^3 \over c^3 m_p m_e \mu_e }
\left(c_s \over c \right)^{1/2} \\
\nonumber &  & \mbox{}  \times {E_0^{4/3}
\rho^{1/6} \over c_s^{5/3} } {(1+z)^2 \over d^2}
\tE^{-1} ~~.
\end{eqnarray}

At $\tE = \tEi$, equations~\ref{numax_exp} and~\ref{fnu_exp} differ
from equations~\ref{numax_powlaw} and~\ref{fnu_powlaw}
by factors of order unity.  This difference
is not worrying since our analytic approximations
are not expected to be particularly
accurate in the transition between the two limiting cases.
Numerical correction factors to the coefficients
of equations~\ref{numax_exp} and~\ref{fnu_exp}
can be derived from numerical integrations.  Such
factors are presented in section~\ref{interp_sec} below.

\goodbreak
\subsubsection{TV Dinner Equations}
We now pause a moment to consolidate our results so far and express
the key equations in terms of fiducial parameter values\footnote{We call
these ``TV dinner equations'' because numerical values for physical
constants have been inserted, so they are ready to use without further
preparation.}.  We begin with equations~\ref{numax_powlaw}
and~\ref{fnu_powlaw}.  These become
\begin{eqnarray} \nonumber
\label{nu_pow_fid}
\lefteqn{ \nuEm = 9.6 \times 10^{12} \times (1+z)^{1/2}
 \left(c_s \over c/\sqrt{3} \right)^{-1/2}} \\
& & \mbox{} \times \left( \xp \over 0.525\right)
\left( \xie \over 0.1 \right)^2 \left(\xib \over 0.1 \right)^{1/2}  \\
\nonumber & \times &
 \left( E_0 / 10^{53} \erg \over \zeta_m^2 / 4 \right)^{1/2}
 \left( \tE \over \day \right)^{-3/2} \Hz 
\end{eqnarray}
and
\begin{eqnarray} \label{fnu_pow_fid}
\lefteqn{ \Fnmxe = 11 \times (1+z)
\left(\phip \over 0.63 \right)  \left(c_s
\over c/\sqrt{3} \right)^{-1/2} } \\
\nonumber & \times & \left(\xib \over 0.1 \right)^{1/2}
 \left(1.3 \over \mu_e \right)
 \left( E_0 / 10^{53} \erg \over \zeta_m^2 / 4 \right) \\
\nonumber & \times & \left( \rho \over 10^{-24} \gram/\cm^3
\right)^{1/2} \left( d \over 4.82 \Gpc \right)^{-2}  \mJy ~~.
\end{eqnarray}
The observed time corresponding to the transition between the power
law and exponential regimes is
\begin{eqnarray} \nonumber
\tEi &=& 12.1 \times (1+z)  \left(c_s \over c/\sqrt{3} \right)^{-8/3}
 \left( E_0 / 10^{53} \erg \over \zeta_m^2 / 4 \right)^{1/3} \\
& \times & \left( \rho \over 10^{-24} \gram/\cm^3 \right)^{-1/3}
 \left( \zeta_m \over 0.1 \right)^{8/3} \days ~~ . 
\end{eqnarray}
Thereafter, the frequency and flux density at the spectral peak are
characterized by equations~\ref{numax_powlaw} and~\ref{fnu_exp}.
Numerical integrations show that modest correction factors $\epsnu
\approx 0.74$ and $\epsF \approx 0.7$ should be applied to these two
equations at late times to compensate for approximations in the
initial conditions (see section~\ref{scaling_sec} below).  These have
been incorporated in the following three equations.

The observed frequency of the spectral peak at the time of the break is
\begin{eqnarray}
\lefteqn{ \nuEmi = {1.7 \times 10^{11} \over 1+z}
\left( {\epsnu \over 0.74} \, {\xp \over 0.525} \right)  \left(c_s
\over c/\sqrt{3} \right)^{7/2} } \\
\nonumber & \times &  \left( \xie \over 0.1 \right)^2
\left(\xib \over 0.1 \right)^{1\over 2} \left( \rho \over 10^{-24}
 \gram/\cm^3 \right)^{1 \over 2} \left( \zeta_m \over 0.1 \right)^{-4} \Hz ~.
\end{eqnarray}

The subsequent evolution is given by
\begin{equation}
\nuEm = \nuEmi \times \left( \tE \over \tEi \right)^{-2}
\end{equation}
and
\begin{eqnarray}
\lefteqn{ \Fnmxe = 0.41 \times  
 \left( {\epsF \over 0.7} \, {\phip \over 0.63} \right)
  \left(c_s \over c/\sqrt{3} \right)^{3/2}
 \left(\xib \over 0.1 \right)^{1/2} }\\
\nonumber &\times &  \left(1.3 \over \mu_e \right)
 \left( E_0 / \zeta_m^2 \over 10^{53} \erg / 4 \right)
 \left( \rho \over 10^{-24} \gram/\cm^3 \right)^{1/2} \\
\nonumber & \times &  (1+z) \left( d \over 4.82 \Gpc \right)^{-2}
  \left( \tE \over \tEi \right)^{-1}
\mJy ~~.
\end{eqnarray}

Finally, for completeness, we include our result for $\nucooE$ from
Rhoads 1999b:
\begin{eqnarray}
\lefteqn{
\nonumber \nucooE = \left[ 5.89 \times 10^{13} \left( \tE / \tEi \right)^{-1/2}
 + 1.34 \times 10^{14} \right] \Hz } \\
& \times &
\left(1 \over 1+z \right)
\left(c_s \over c/\sqrt{3} \right)^{17/6}
\left( \xiB \over 0.1\right)^{-3/2} \\
\nonumber & \times & \left( \rho \cdot \cm^{3} \over 10^{-24} \gram 
\right)^{-5/6} \left( E_0 / 10^{53} \erg \over \zeta_m^2 / 4 \right)^{-2/3}
\left( \zeta_m \over 0.1 \right)^{-4/3} ~~.
\end{eqnarray}
Note that this equation already interpolates over the break time $\tEi$; 
the interpolation was derived in the fashion suggested in
section~\ref{interp_sec} below.

\subsubsection{Putting the Pieces Together} \label{scaling_sec}
\begin{figure*}[htb] 
\epsfxsize=0.9\hsize \epsfbox{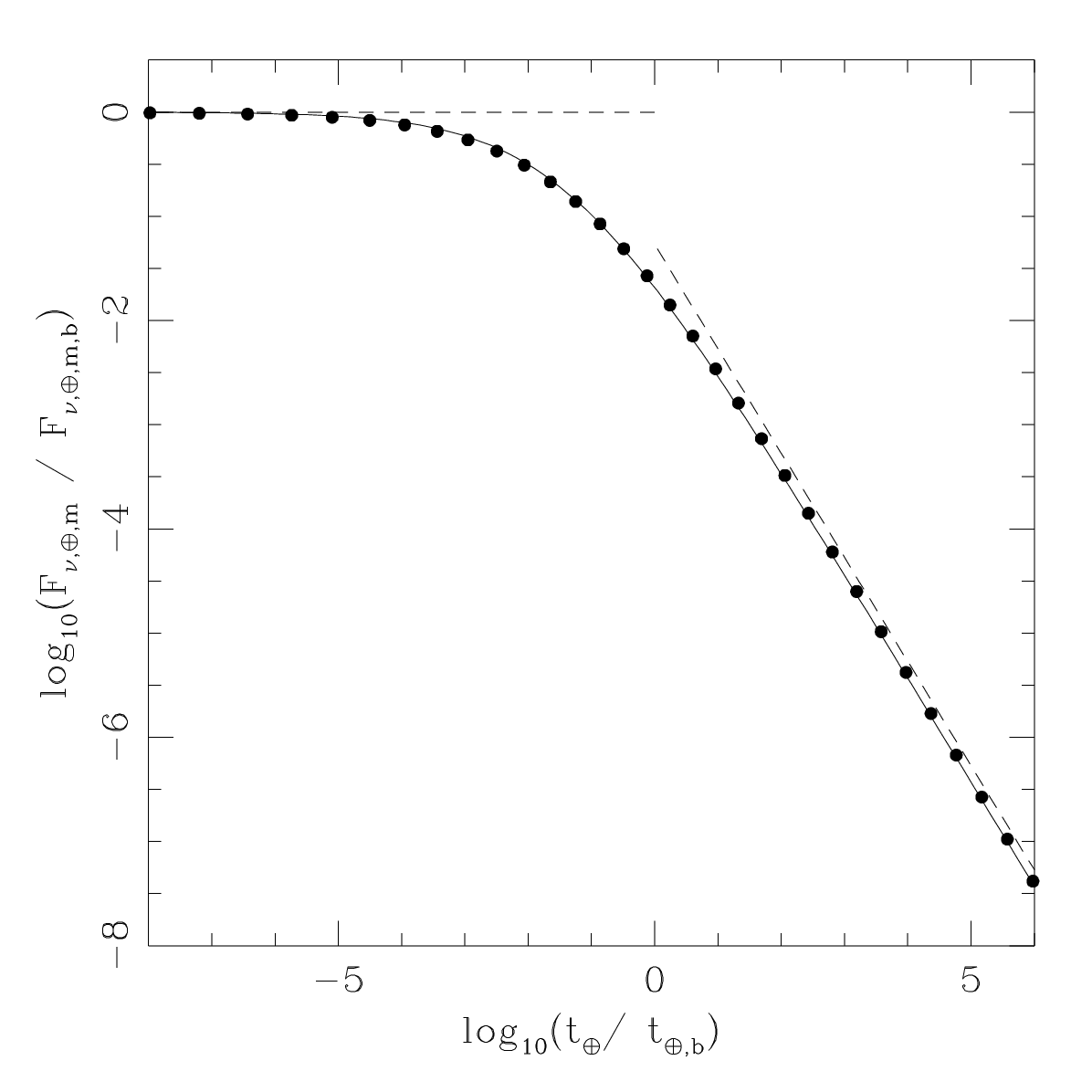}
\vspace{10pt}
\caption{Dimensionless peak flux as a function of dimensionless
observer-frame time.  Points show the results of numerical
integrations.  Dashed lines show the analytic asymptotic forms from
equations~\ref{fnu_powlaw} and \ref{fnu_exp}, scaled to fiducial
values as described in section~\ref{scaling_sec}.  The late-time flux
density is below the prediction of equation~\ref{fnu_exp} by a factor
of $\sim 0.7$.  This stems from approximations in the exponential
regime initial conditions (equations~\ref{initc_first} to
\ref{initc_last}), which are derived by applying the power law regime
results beyond their range of strict validity.  The solid curve shows
an empirical interpolation between the early and late-time analytic
forms, incorporating a factor $0.7$ correction to the late-time asymptotic
form  (see section~\ref{interp_sec}).}
\label{fnu_fig}
\end{figure*}

An accurate description of the behavior in the transition between the
power law and exponential regimes can be obtained numerically.  We
first note that there is a single characteristic observed time $\tEi$
given by equation~\ref{initc_tE} and flux level $\Fnmxei \equiv
\Fnmxe(\tE \ll \tEi)$ given by equation~\ref{fnu_powlaw}.
If we use these as our basic time
and flux units, and denote the observed time and peak flux scaled to
these units as $\hattE$ and $\hatFnmxe$, there is a unique
$\hatFnmxe(\hattE)$ relation.  This is plotted in figure~\ref{fnu_fig}.

\begin{figure*}[htb] 
\epsfxsize=0.9\hsize \epsfbox{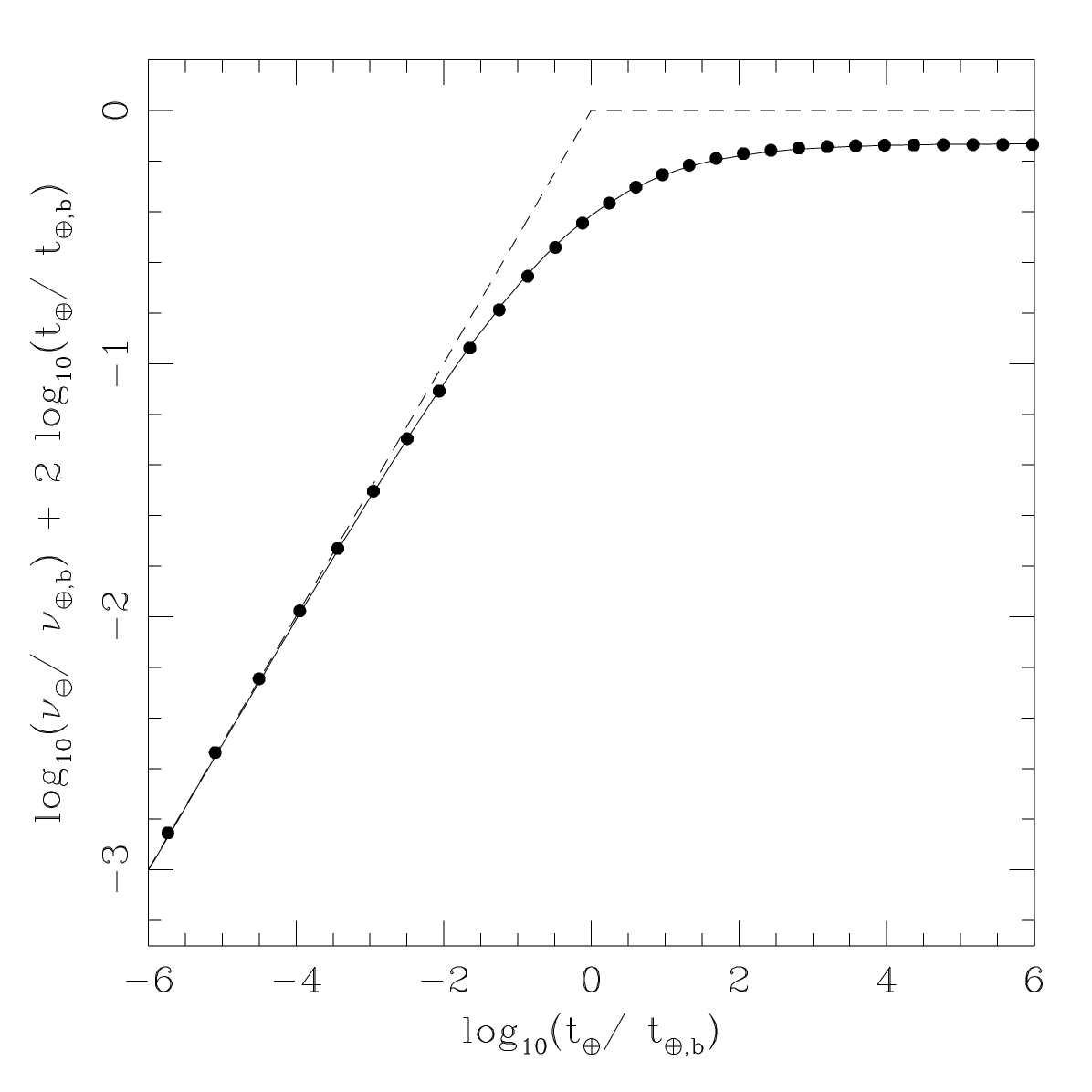}
\vspace{10pt}
\caption[nonoptionalargument]{Dimensionless frequency at which the
spectrum peaks as a function of dimensionless observer-frame time.
The $y$-axis shows $\log_{10}(\nuEm \tE^2)$, so that a $\tE^{-2}$
decay of peak flux density with time appears as a horizontal line.  As
in figure~\ref{fnu_fig}, points show the results of numerical
integrations; dashed lines show analytic asymptotic forms (here from
equations~\ref{numax_powlaw} and \ref{numax_exp}), and the solid line
is an empirical interpolation.  The late-time asymptotic frequency
given by equations~\ref{numax_exp} is seen to be too large by a factor
$\sim 1.35$.  This is due to the approximate initial
conditions used to derive equation~\ref{numax_exp}, and a correction
factor has been applied in deriving the interpolated curve (see
section~\ref{interp_sec}).  }
\label{numax_fig}
\end{figure*}

Similarly, we can define the characteristic frequency $\nuEmi$ in the
problem to be given by equation~\ref{numax_powlaw} evaluated at
$\tEi$, and $\hatnuEm$ to be the frequency scaled by this value.  Then
we can again obtain a unique relation $\hatnuEm(\hattE)$, which is
shown in figure~\ref{numax_fig}.

At late times, the numerical integrations yield a flux density that is
a factor $\epsF \approx 0.7$ smaller than in
equation~\ref{fnu_exp}, and a frequency of peak emission that is
a factor $\epsnu \approx 0.74$ smaller than in
equation~\ref{numax_exp}.  This is presumably due to the approximate
initial conditions used for the exponential regime evolution.  These
initial conditions are obtained by applying an asymptotic
approximation outside its range of validity, and it should not be
surprising if this procedure introduces some error.  We suggest below
that this error may be corrected empirically.

To obtain predictions for a given set of model parameters from these
dimensionless curves, we need only (1) determine numerically the
values of $\tEi$, $\Fnmxei$, and $\nuEmi$; and (2) determine the time
interval over which our assumption $ 1/\Gamma_0 \la f \la \Gamma_0$
remains valid.  The early behavior, before the ejecta accrete a
dynamically important amount of ambient medium (i.e., $f <
1/\Gamma_0$) is unlikely to be observed at long wavelengths,
since it is over within a
fraction of a second for reasonable burst parameters.  We therefore
consider only the end condition, $f \approx \Gamma_0$.  This happens
at $\tE = \tEf$, where
\begin{equation}
\tEf \approx \Gamma_0 \tEi / \finit \approx {8\over 25} \left(c_s
\over c \right)^2 {1 \over \zeta_m^2 } \tEi ~~.
\end{equation}
At later times, our assumptions that $\Gamma \ga 2$ and $\beta \approx
1$ break down, and the behavior of the fireball changes again.  Such
changes may be relevant to the radio behavior of gamma ray burst
afterglows, but we will not consider them here.

\subsubsection{Empirical Interpolations} \label{interp_sec}
To obtain a readily calculated burst behavior around time $\tEi$, we
can interpolate between the asymptotic behaviors for earlier and later
times.  We do this first for $\Fnmxe$ and then for $\nuEm$.  We use
interpolants of the form $g = ( g_1^{-\kappa} + \epsilon g_2^{-\kappa}
)^{-1/\kappa}$, where $g_1$ and $\epsilon g_2$ represent limiting
behaviors of an arbitrary function $g$ for early and late times.  The
exponent $\kappa$ determines the smoothness of the transition between
the limiting behaviors.  The scalar $\epsilon$ is introduced so that the
numerically derived correction factors to the late-time asymptotic results
can be applied.

For $\Fnmxe$, the asymptotic behaviors are $\Fnmxe$ constant and
$\Fnmxe\sim \hattE^{-1}$.  We work with the scaled quantities defined
in section~\ref{scaling_sec}, so that the break between the two
asymptotic behaviors is expected for $\log(\hattE) \sim 0$.  We set
$g_1 = \Fnmxei$.  We use equation~\ref{fnu_exp} for $g_2$, and set the
correction factor $\epsilon = 0.7$.  Finally, we choose $\kappa =
0.4$.  The resulting interpolation is plotted atop the numerical
integration results in figure~\ref{fnu_fig}.

The asymptotic behaviors of $\nuEm$ are $\propto \tE^{-3/2}$ and
$\propto \tE^{-2}$.  In this case, we have taken $g_1$ from
equation~\ref{numax_powlaw}.  For $g_2$, we take
equation~\ref{numax_exp}, and set $\epsilon = 0.74$.  Here we find
$\kappa = 5/6$ works well.  This interpolation is shown in
figure~\ref{numax_fig}.

\subsubsection{Light Curves: Optically thin case} \label{lcurv_othin}
The afterglow light curve at fixed observed frequency is obtained by
combining the predicted behavior of $\nuEm$ and $\Fnmxe$ with the
spectrum for a truncated power law electron energy distribution.  We
use the analytic results of section~\ref{analytic_othin}.
Then we find four
generic behaviors, depending on whether the frequency is above or
below $\nuEm$ and whether the time is earlier or later than $\tEi$.
These are
\begin{equation} \label{lcurv_wax}
\FE \propto \left\{
\begin{array}{lll}
\tE^{1/2} & \tE < \tEi ~; & \nuE < \nuEm(\tE) \\
\tE^{-3(p-1)/4} & \tE < \tEi ~; & \nuE > \nuEm(\tE) \\
\tE^{-1/3} & \tE > \tEi ~; & \nuE < \nuEm(\tE) \\
\tE^{-p} & \tE > \tEi ~; & \nuE > \nuEm(\tE)
\end{array}
\right.
\end{equation}
Here $p$ is the electron energy spectrum slope and $\alpha = (p-1) /
2$ is the high frequency spectral slope, as usual.  Note particularly
how steep the light curve becomes for $\tE > \tEi$ and $\nuE >
\nuEm(\tE)$.

\begin{figure*}[htb] 
\epsfxsize=0.9\hsize \epsfbox{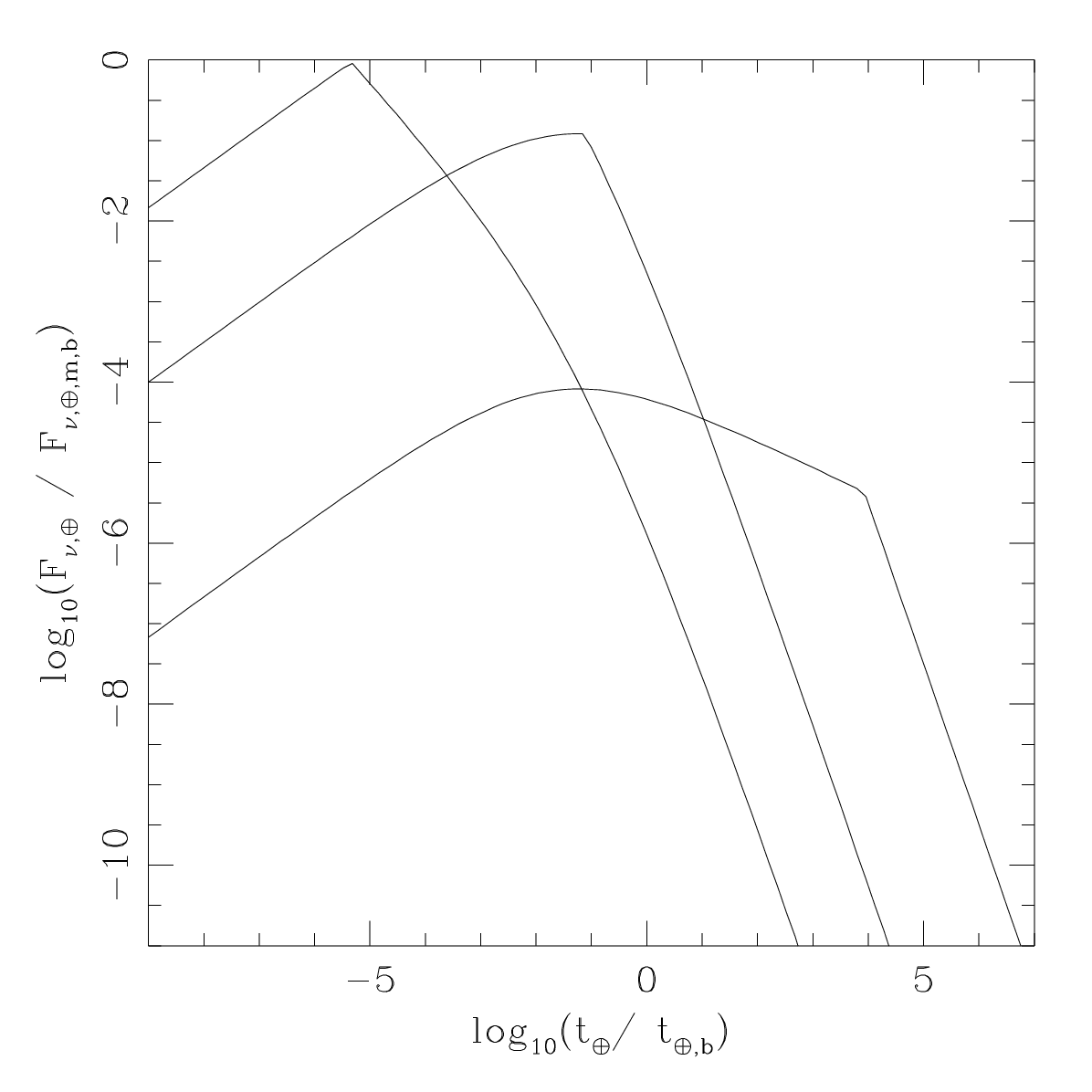}
\vspace{10pt}
\caption{Sample light curves in dimensionless units.
The curves have been derived by combining a broken power law spectrum
(with electron energy distribution slope $p=2$) with the interpolated,
dimensionless forms of the $\Fnmxe(\tE)$ and $\nuEm(\tE)$ relations
(see section~\ref{interp_sec} for details).  The light curves
correspond to frequencies $10^{8} \nuEmi$, $10^{1.5}\nuEmi$, and
$10^{-8}\nuEmi$, in order of decreasing flux density at the earliest
time plotted.  All the asymptotic behaviors described in
equations~\ref{lcurv_wax} are exhibited here, though the additional
cooling break (equation~\ref{lcurv_cool}) is omitted.  The
transition between the behaviors for $\tE \ll \tEi$ and  $\tE \gg
\tEi$ is rather gradual.  The other transition, between $\tE \ll
\tEm(\nuE)$ and  $\tE \gg \tEm(\nuE)$, is artificially sharp in these
plots because the adopted spectrum has a discontinuous slope at
$\nuEm$.  A more detailed treatment of the spectrum would smooth out
this transition also.
}
\label{lcurv_fig}
\end{figure*}

Three representative light curves are shown in figure~\ref{lcurv_fig}.
These have been derived by combining the empirically interpolated
$\hatnuEm$ and $\hatFnmxe$ curves with the broken power law spectrum.
Note that the rollover at the beaming transition ($\log(\hattE) \sim 0$) is
rather slow, so that observed behavior will be intermediate between
the asymptotic power laws of equations~\ref{lcurv_wax} for a
considerable time.  This slow rollover is in part due to the compound
nature of the break.  The light curve decay begins to accelerate as soon
as we can ``see'' the edge of the jet, when $\Gamma < 1/\zeta_m$.  The
additional steepening when dynamical effects of beaming become
important occurs slightly later, when $\Gamma < \Gami \sim 0.23 /
\zeta_m$ (cf. equation~\ref{initc_Gam}) (cf. Panaitescu \& \mesz 1999
for additional discussion of this point).

Equation~\ref{lcurv_wax} assumes $\nuabsE < \nuE < \nucooE$
throughout.  (Here $\nuabsE$ is the self-absorption frequency,
measured in the observer's frame.)  If we now include the cooling
break, we obtain the additional light curve behaviors (derived in
Rhoads 1999b)
\begin{equation} \label{lcurv_cool}
\FE \propto \left\{
\begin{array}{lll}
\tE^{1/2 - 3 p / 4} & \tE < \tEi ~; & \nuE > \nucooE(\tE) \\
\tE^{-p} & \tE > \tEi ~; & \nuE > \nucooE(\tE)
\end{array}
\right.
\end{equation}
where we have also assumed that $\nuEm < \nucooE$.
These behaviors are not shown in figure~\ref{lcurv_fig}, but were used
to fit the light curve of GRB 970508 with beamed afterglow models
(Rhoads 1999b).

\subsubsection{Light Curves: Optically thin case without sideways expansion}
It remains possible to constrain gamma ray burst beaming by looking
for light curve breaks even in the case where lateral expansion of the
evolving burst remnant is unimportant.  This corresponds to dropping
our fourth assumption from section~\ref{dynamics}.

In this case, the dynamical evolution follows power law behavior
(section~\ref{pl_dyn_analytic}) throughout, but the emergent radiation is
diluted relative to the spherical case by a factor $\Gamma^2
\zeta_m^2$ once $\Gamma < 1/\zeta_m$.  For an adiabatic
evolution, we have $\Gamma^2 \propto \tE^{-3/4}$.  The power law exponents
for the light curve in this regime become $-1/4$ for $\nuabsE < \nu <
\nuEm$, $-3 p / 4$ for $\nuEm < \nu < \nucooE$, and $-1/4 - 3 p / 4$
for $\nucooE < \nu$.

The most plausible mechanism for quenching lateral expansion is
a strongly radiative remnant, so it is more relevant to examine
this regime with radiative dynamics.  Then $\Gamma^2 \propto
\tE^{-6/7}$, and all the light curve exponents decrease by an
additional $3/28$, becoming $-5/14$ for $\nuabsE < \nu < \nuEm$,
$-3/28 - 3 p / 4$ for $\nuEm < \nu < \nucooE$, and $-5/14 - 3 p / 4$
for $\nucooE < \nu$.

While these slope changes are less dramatic than those in 
section~\ref{lcurv_othin}, they would be strong
enough to detect in afterglow light curves with reasonably large time
coverage and good photometric accuracy.

\subsection{Optically Thick Case}
We now consider briefly the electron energy distribution model of \bp\
\& Rhoads (1993).  This model differs from that of the preceding
sections in a few ways.  First, the electron power law index was fixed
at $p=2$ to avoid strong divergences in the total energy density in
electrons.  Second, the minimum electron energy $\eemin$ was taken to
be sufficiently small that emission from electrons with $\ee = \eemin$
was always in the optically thick regime.  Under these circumstances,
there is a single break in the electron energy spectrum at the
frequency corresponding to optical depth $\tau = 0.35$ (cf. Pacholczyk
1970), with spectral slope $\nu^{5/2}$ below the break and
$\nu^{-1/2}$ above.  The magnetic field behavior is the same in this
model and more recent ones.

Combining this electron behavior with the power law regime dynamical
model reproduces the scalings from \bp\ \& Rhoads (1993), namely
\begin{equation}
\nuEm \propto E_0^{1/3} \rho^{1/3} \tE^{-2/3}
\end{equation}
and
\begin{equation}
\FEm \propto E_0^{7/8} \rho^{1/8} \nuEm^{5/8}
 \propto E_0^{13/12} \rho^{1/3} \tE^{-5/12} ~~.
\end{equation}

If we use instead the exponential regime dynamical
model, we find
\begin{equation}
\nuEm \propto \tE^{-1} \qquad \hbox{and} \qquad
\FEm \propto \tE^{-3/2} ~~.
\label{bpjer_exp_ev}
\end{equation}

Readers interested in the precise numerical coefficients for these
relations are referred to \bp\ \& Rhoads (1993) for the spherical
case.  For the beamed case, numerical results may be found by applying the 
\bp\ \& Rhoads (1993) results at the transition between the power law
and exponential regimes, and continuing the evolution using
equations~\ref{bpjer_exp_ev}.

The light curve for this electron model then becomes
\begin{equation}
\FE \propto \left\{
\begin{array}{lll}
\tE^{5/4} & \tE < \tEi ~; & \nuE < \nuEm(\tE) \\
\tE^{-3/4} & \tE < \tEi ~; & \nuE > \nuEm(\tE) \\
\tE^{1} & \tE > \tEi ~; & \nuE < \nuEm(\tE) \\
\tE^{-2} & \tE > \tEi ~; & \nuE > \nuEm(\tE)
\end{array}
\right.
\end{equation}
The behavior here at frequencies $\nuE > \nuEm(\tE)$ is the same as in
equations~\ref{lcurv_wax} with $p=2$.  However, $\nuEm$ has
a different meaning in the two models.

From these results, we see that the substantial changes in the
observable behavior of a beamed burst are not dependent on the precise
nature of the electron energy distribution.

\section{Discussion} \label{discuss}
We now put mathematics aside to recapitulate our results and to
discuss their implications for the interpretation of afterglow
observations.

We have shown that the dynamics of a gamma ray burst remnant change
qualitatively when the remnant's Lorentz factor $\Gamma$  drops
below the reciprocal opening angle $1/\zeta_m$ of the ejecta.  Before
this time, the Lorentz factor behaves as a power law in radius.
Afterwards, the Lorentz factor decays exponentially with radius.  The
change occurs because lateral expansion of the ejecta cloud increases
the rate at which additional material is accreted.  Such lateral
expansion is prohibited by symmetry in the spherical case.

When the remnant enters this ``exponential regime,'' the relation
between the observed spectrum and the observed light curve changes.
Inferences about the electron energy spectrum in afterglows come from
the light curve decay rate and spectral slope.  The general agreement
between the two methods has been taken as a confirmation of the
(spherically symmetric) fireball model (Wijers, Rees, \& \mesz\ 1997;
Waxman 1997a).

The light curve decline at frequencies above the spectral peak becomes
very steep ($\tE^{-p}$, where $p$ is the index of the electron energy
spectrum) once the burst dynamics enter the exponential regime.
Reconciling this relation with the observed decays ($-1 \ga d
\log(\FE) / d \log(\tE) \ga -1.5$) would require an extremely flat
electron energy spectrum, and consequently a very blue spectral energy
distribution.  This was not seen in early observed spectral energy
distributions (see Wijers et al 1997 for GRB 970228; Reichart 1997 and
Sokolov et al 1997 for GRB 970508; and Reichart 1998 for GRB 971214).
We infer that GRBs 970228, 970508, and 971214 were probably not in the
exponential regime during their observed optical afterglows.  
GRB 971227 provides a possible, though dubious, counterexample.  There
is one image suggesting a counterpart (magnitude $R\approx 19.5$) on
December 27.91 (Castro-Tirado et al 1997).  Later images show no
corresponding source,  requiring a decay at least as fast as
$\tE^{-2.5}$ (Djorgovski et al 1998).  This is consistent with a
$\tE^{-p}$ decay for typical values of $p$.  However, this explanation
remains speculative, since there is no second image confirming the proposed
counterpart.

Subsequent afterglows have provided a more hopeful picture for
practical application of beaming models.  In particular, GRB 990123
shows a break that is quite possibly due to beaming (e.g.,
Castro-Tirado et al 1999; Kulkarni et al 1999), and comparison of the
spectral slope and decay slope for GRB 980519 gives better agreement
for beamed than for spherical regime models (Sari et al 1999).

In the case of GRB 970508, we can place a stringent limit on the
beaming angle in the context of our model.  The optical light curve
extends to $\sim 100 \days$ after the burst and does not depart
drastically from a single power law after day 2 (Pedersen et al 1998);
thus, no transition to the exponential regime occurred during this
time.  As already noted, the spectral slope and light curve law decay
rate are in fair agreement for the spherical case, and poor agreement
for the beamed case.  The radio light curve furnishes the last
critical ingredient.  Goodman (1997) pointed out that diffractive
scintillation by the Galactic interstellar medium is expected in early
time radio data, and that this scintillation will stop when the
afterglow passes a critical angular size.  By comparing this
characteristic size with the time required for the scintillations to
die out, one can measure the burst's expansion rate.  This test has
been applied (Frail et al 1997; Waxman, Kulkarni, \& Frail 1998)
and shows that $\Gamma \la 2$ at $\tE \sim 14 \days$.
Thus, no power-law break to faster decline is observed at $\Gamma \ga
2$, and we infer that GRB 970508 was effectively unbeamed ($\zeta_m
\ga 1/2$).
This rough derivation is borne out by detailed fitting of beamed
afterglow models to the GRB 970508 light curve, which yields the same
beaming limit $\zeta_m \ga 1/2$ radian (Rhoads 1999b).

This conclusion, combined with the GRB 970508 redshift limit $z \ge
0.835$ (Metzger et al 1997), immediately implies a minimum energy for
the burst.  This burst was detected as BATSE trigger 6225, and the
total BATSE fluence was $(3.1 \pm 0.2) \times 10^{-6} \erg / \cm^{-2}$
over the range $20$--$1000 \keV$ (Kouveliotou et al 1997).  The gamma
ray emission alone therefore implies $E_0 \ga 4.7 \times 10^{51}
(\Omega / 4 \pi) \erg \ga 3 \times 10^{50} \erg$.  Here we have based
the luminosity distance on an $\Omega = 0.2$, $\Lambda = 0$, $H_0 = 70
\km/\sec/\Mpc$ cosmology, and applied the beaming angle limit $\zeta_m
\ga 0.5$ radian in the second inequality.  This conclusion is of
course model-dependent and might change if our assumptions about the
blast wave physics or beaming geometry are badly wrong.  We will
discuss possible ways to reduce the energy requirements of GRB 970508
while retaining consistency with the afterglow data in
section~\ref{min_en} below.



If the beaming angle $\zeta_m$ is substantially variable from burst to
burst, it is possible that some bursts enter the rapid decay phase
before the spectral peak passes through optical wavelengths.  Present
data suggests that this is indeed the case; GRB 980519 is best fit by
assuming exponential regime behavior (Sari et al 1999), while GRB
990123 appears to be a transition case with a break observed in the
optical light curve (e.g. Castro-Tirado et al 1999, Kulkarni et al
1999, Sari et al 1999).  The resulting rapid decay could then explain
some of the optical non-detections of well studied GRBs such as 970828
(Groot et al 1997).  Alternatively, for characteristic beaming angles
$1 \gg \zeta_m \ga 0.1$, we would expect beaming to become dynamically
important between the time of peak optical and radio afterglow.  This
would then help explain the paucity of radio afterglows, which unlike
optical afterglows cannot be hidden by dust in the burster's
environment.  There is some evidence that the radio emission involves
a different process, or at least a different electron population, from
the optical and X-ray afterglows: The peak flux density in GRB 970508
did not follow a single power law with wavelength as it ought to under
the simplest fireball models (Katz \& Piran 1997b).

The transition in light curve behavior at $\Gamma \sim 1/\zeta_m$ is
also important for ``blind'' afterglow searches.  Such searches would look
for afterglows not associated with detected gamma ray emission.  A
much higher event rate for afterglows than for bursts is a natural
consequence of beamed fireball models, since the afterglow emission
peaks at lower bulk Lorentz factors than the gamma ray emission does.
Comparison of event rates at different wavelengths can therefore constrain
the ratio of beaming angles at those wavelengths (Rhoads 1997a).
However, we will only see the afterglow if either (a) we are within
angle $\zeta_m$ of the burst's symmetry axis, and therefore could also
see the gamma ray burst, or (b) the Lorentz factor has decayed to
$\Gamma < 1/\zeta_m$ and the afterglow light curve has entered its
steep decay phase.  We have already argued that GRB 970228, GRB
970508, and GRB 971214 were not in this steep decay phase based on the
comparison of light curves and spectral slopes.  It follows that if blind
afterglow searches find a population of afterglows not associated with
observed gamma rays, those afterglows will exhibit a steeper light
curve decay than did the 1997 afterglows.  The efficiency
for detecting such rapidly fading ``orphan'' afterglows will be substantially
lower than the efficiency estimated from direct comparison with
spherical-regime afterglows.

Other models of beamed gamma ray bursts are possible.  In particular,
we have assumed a ``hard-edged'' jet, where the mass and energy
emitted per unit solid angle are constant at small angles and drop to
zero as a step function at large angles.  Profiles in which these
quantities decrease smoothly to zero may be more realistic.  Whether
these differ importantly from the model presented here depends on
whether most of the energy is emitted into a central core whose
properties vary slowly across the core.  Layered jet models in which
most of the kinetic energy from the fireball is carried by material
with a low Lorentz factor can have substantially different afterglow
light curves from either the spherically symmetric case or the
hard-edged jet case.  This is because the afterglow emission can be
dominated by outer layers where the initial Lorentz factor is high
enough to yield optical emission during ejecta deceleration, but
insufficient to yield gamma rays.  The afterglow is thereby
effectively decoupled from the gamma ray emission, and it becomes
harder to predict one from the other.  Such models have been explored
by several groups (e.g., \mesz \& Rees 1997b; \mesz, Rees, \& Wijers
1997; \bp\ 1997).  A similar decoupling of the gamma-ray and afterglow
properties can be produced in the spherical case by allowing inner
shells of lower Lorentz factor and larger total mass and energy to
follow the initial high-$\Gamma$ ejecta (Rees \& \mesz\ 1998).

It is possible to approximate the afterglow from a layered jet by a
superposition of hard-edged jets.  For this to be reasonably accurate,
the outer layers should have Lorentz factors substantially below those
of the inner layers,  and opening angles and energies substantially
above those of the inner layers.


\bigskip\bigskip
\goodbreak 
\subsection{Energy Requirements for GRB 970508} \label{min_en}
We now consider how our model will change if we vary some of the basic
assumptions.  Our primary concern is to determine whether the minimum
energy required to power the GRB 970508 afterglow can be reduced
substantially below the requirements derived from a spherical
adiabatic fireball model expanding into a homogeneous medium.  We will
therefore sometimes err on the side of extreme model assumptions
chosen to minimize the energy needs.  In order to declare a model
consistent with the data, we require that either (1) there be no break
in the light curve or spectrum around $\Gamma \sim 1/\zeta_m$, or (2)
the break occurs early (before $\tE \sim 2 \days$) and the late
time light curve shows a slow decline even for spectral slopes as red
as those observed.

The first requirement is physically implausible.
Even in the absence of the dynamical effects reported above, so long
as the afterglow is from relativistically moving and decelerating
material, its flux will scale with an extra factor $\Gamma^2$ once
$\Gamma \ga 1/\zeta_m$.  Since $\Gamma$ decreases with time, a break
is generally expected, though perhaps it could be avoided with
sufficient fine-tuning of the model.  

The second possibility is more interesting.  It requires us to
construct a model where factors besides beaming contribute relatively
little to the decay of $\Fnmxe$ with $\tE$, or where the observed
spectrum does not directly tell us about the electron energy
distribution.  A burst expanding into a cavity (such that $\rho$
increases with $r$) might give a slow decay, while a sufficiently
large dust column density would give a red spectrum despite a flat
electron energy distribution (cf. Reichart 1997, 1998).  However, both
would require some degree of fine tuning.  The dust-reddened spectra
would deviate measurably from pure power-laws given good enough
data, but the present data are probably equally consistent with both
pure and reddened power law spectra.  Certainly such reddened beaming models
would imply little correlation between observed spectral slopes and
light curve decays, since the dust column density could vary wildly
from burst to burst.  This hypothesis is somewhat ad hoc, but is
consistent with present data and is supported by other circumstantial
evidence linking GRBs to dust and star forming regions (e.g. Groot et
al 1997; \bp\ 1998).  At present, then, it appears the most viable way
of reconciling beamed fireball scenarios with the 1997 afterglow data.

We now discuss a few variations of fireball models in greater detail.

\subsubsection{Radiative case}
We first consider the behavior of a radiative regime fireball.  In this
regime, the internal energy of the fireball is low, since it is
converted to photons and radiated away.  The largest implications for
beaming are when the internal energy density is so low that $c_s \ll
c$.  In this case, the lateral expansion that leads to the exponential
regime of burst remnant evolution in the adiabatic case is
unimportantly small.  We  assume this low sound speed through much
of the following discussion.

We assume that energy in magnetic fields and protons is transferred to
electrons in the burst remnant on a remnant crossing time ($\sim
t_{co}$).  The electrons are assumed to maintain a power law energy
spectrum, with a large $\eemax$ whose precise value is determined by
the requirement that the burst radiate its internal energy
efficiently.  Under these circumstances, the Lorentz factor scales as
$\Gamma \propto r^{-3}$ and the comoving frame internal energy $\Eint$
of the remnant follows the evolution
\begin{equation}
{d \Eint \over d r} \sim \Gamma \rho c^2 \pi \zeta_m^2 r^2  - {\Eint \over
t_{co}} { d t_{co} \over d r }
 \sim { \pi \rho c^2 \zeta_m^2 (\Gamma r^3) \over r } - { 4 \Eint \over r } ~~.
\end{equation}
This admits a solution of the form $\Eint \sim (\pi \zeta_m^2 \rho c^2
(\Gamma r^3) - c_2 r^{-4} )/4$ where $c_2$ is a constant of
integration.  At late times, we throw away the $c_2 r^{-4}$ term,
which becomes negligible.  The result then becomes
\begin{equation}
\Eint \approx (\Gamma r^3) \pi \zeta_m^2 \rho c^2 / 4
\end{equation}
which is constant since $\Gamma \propto r^{-3}$ in this regime.
If the sound speed becomes negligibly small at some  point in the
burst remnant evolution, then the volume of the shell scales as $V
\propto r^2$ thereafter.  The magnetic field then scales as $B \propto
r^{-1}$, based on constant $\Eint$.  The observed peak frequency
scales as $\nuEm \propto \Gamma^3 B \propto r^{-10} \propto \tE^{-10/7}$.

The power radiated is simply 
$\sim \Gamma \times (\Gamma - 1) c^2 \times \pi \zeta_m^2 r^2 c \rho $
in the observer's frame, but this is dominated by emission from
electrons at $\eemax$, which we have not calculated.  The peak in $\FE$
must be estimated as before.  We have total comoving frame energy
$\sim \xie \Eint$ in electrons at $\ee \sim \eemin$, which is radiated
over the comoving cooling time $t_s \sim 1 / (\Gamma B^2) \sim V /
(\Gamma \Eint) \sim r^5$.  In the observer frame, this gives total
power output $\sim \Gamma^2 \Eint / t_s$, accounting for factors of
$\Gamma$ from the Lorentz boost to photon energies and for the
transformation between $t_{co}$ and $\tE$.  The frequency range
containing this power scales as $\Delta \nuE \propto \nuEm$.  So, in
the spherical case, $\Fnmxe \propto \Gamma^2 \Eint / (t_s \nuEm)
\propto r^{-1} \propto \tE^{-1/7}$.
If we now allow for beaming, we introduce another factor of
$\Omega_\gamma^{-1} = \Gamma^2 / \pi$ and obtain $\Fnmxe \propto
r^{-7} \propto \tE^{-1}$.

Finally, putting in the spectral shape for fixed $\nuE >
\nuEm$, we find that $\FE \propto \tE^{-(5 p + 2)/7}$.  Thus, this
radiative regime model yields scalings fairly similar to our canonical
adiabatic model.  In particular, the late time light curve again shows
a steep decline.  While the assumptions made here may not be fully
self-consistent, allowing $c_s \sim c$ would likely further steepen
this decline.

This result suggests that the GRB 970508 data cannot easily be
reconciled with a beamed radiative afterglow model.

\subsubsection{Beamed burst, isotropic afterglow}
Gamma ray bursters may give rise to both fast and slow ejecta, where
``slow'' here means too slow to cause gamma ray emission.  In this
case, the optical and $\gamma$-ray properties of the event may be
effectively decoupled if the slow wind contains most of the energy.

Suppose the gamma ray burst is caused by a small amount of extremely
relativistic ejecta, while the afterglow is caused primarily by a
greater mass of material with low $\Gamma_0$.  The afterglow light
curve places almost no direct constraint on the isotropy of the first
(high-$\Gamma_0$) material.  However, the low-$\Gamma$ material must
be reasonably isotropic to avoid a visible break in the light curve at
late time.  To explain a peak optical flux of $\sim 30 \microJy$ in
the optical, we need total energy 
\begin{eqnarray} \nonumber
\lefteqn{ E_0  = 1.5 \times 10^{50} \erg \times \left( {0.63 \over \phip}
{\mu_e \over 1.3} \right)
\left(c_s \over c/\sqrt{3} \right)^{1/2}
\left( \xib \over 0.1\right)^{-1/2} } \\
& \times & \left( \rho \cdot \cm^{3} \over 10^{-24} \gram   \right)^{-1/2}
\left(1.835 \over 1+z \right) \left( d \over 4.82 \Gpc \right)^2
\label{minen_th}
\end{eqnarray}
(see equations~\ref{fnu_powlaw} and \ref{fnu_pow_fid}),
where we have assumed isotropy and where $4.82 \Gpc$ is the luminosity
distance corresponding to $z=0.835$ for cosmological parameters
$\Omega = 0.2$, $\Lambda=0$, and $H_0 = 70 \km/\sec /\Mpc$ (cf.\
equation~\ref{fnu_powlaw}).  We compare this to the optical fluence of
the burst, which we estimate as
\begin{equation}
Q_{opt} = \int_0^\infty \int_0^{\nu_{max}} \FE d\nu d \tE~~.
\end{equation}
Inserting our broken power law spectrum and the dependence of $\nuEm$
on $\tE$ yields 
\begin{equation}
Q_{opt} \approx \left( {2\over 3} + {4 \over p - 7/3} \right) \Fnmxe \nu_1
\tEm(\nu_1)  \left( \nu_{max} \over \nu_1 \right)^{1/3} ~~.
\label{flu_int}
\end{equation}
Here $\nu_1$ is an arbitrary reference frequency, and $\tEm(\nu_1)$ is
defined as the moment when $\nuEm = \nu_1$.  Setting $\nu_1 = 6 \times
10^{14} \Hz$ (corresponding to wavelength $0.5 \micron$), $\Fnmxe = 30
\microJy$, $\tEm(\nu_1)= 2$ days, and $p \approx 2.85$ (corresponding
to a $\tE^{-1.4}$ light curve) yields $Q_{opt} = 3.0 \times 10^{-7}
(\nu_{max}/\nu_1)^{1/3} \erg/\cm^2$.  (A similar calculation using
$p=2.2$ and accounting for the additional break at the cooling
frequency yields a similar fluence, $3.8\times 10^{-7} \erg/\cm^2$, for
$\nu_{max}=\nu_1 = 6 \times 10^{14} \Hz$.  This is the value used in
Rhoads 1999b.)
Taking luminosity distance $4.82 \Gpc$ and considering only optical
and longer wavelength afterglow, the smaller optical fluence estimate
implies $E = 4.5 \times 10^{50} (\Omega/4 \pi)
\erg \ga 2.8 \times 10^{49} \erg$.  We have applied our beaming
limit, $\zeta_m \ga 0.5$ radian, to derive the lower limit here.
If we take $\nu_{max}$ corresponding to the soft X-ray afterglow, the energy
rises by another factor of $\sim 10$.  These fluence-based energy
needs are dangerously close to exceeding the energy requirements from
equation~\ref{minen_th}.  Since the latter equation is based on an
energy-conserving model, this comparison shows that $\xib$ must be
substantially below $1$ and/or the density substantially below
$10^{-24} \gram/\cm^3$ if the model is to be self-consistent.
Otherwise, the total energy radiated is comparable to or greater than
the total energy available.  Reassuringly, the density and magnetic
energy fraction found by Wijers \& Galama (1998) are roughly
consistent with these requirements.  This consistency check could be
refined by replacing equation~\ref{flu_int} with a more
detailed fluence calculation.

\subsubsection{Layered jet models}
In this class of models, considered (for example) by \mesz, Rees, \&
Wijers (1997) and Panaitescu, \mesz, \& Rees (1998), the material
dominating the emission changes continuously through a range of
initial Lorentz factors.  We can approximate such models as a
superposition of many ``hard-edged'' jet models.  We have tried
developing such models while minimizing the energy requirements.  To
do this, we build a sequence of adiabatic hard-edged jets, enforcing
either the condition $\nuEm = \nuE$ or the condition $\tEi = \tE$
throughout the afterglow evolution, and then adjusting the input
energy requirement to match the observed light curve.  (Here $\nuE$
denotes the fixed frequency at which our data was taken.)  A
preliminary exploration of such models has not yielded any drastic
reduction in energy requirements.  A more thorough study may be needed
to make this conclusion firm.

\subsubsection{Inhomogeneous environments}
The predictions of fireball models change somewhat if the ambient
medium is not uniform.  To date, investigations of variable density
environments have concentrated on density laws $\rho \propto
r^{-g}$ (Vietri 1997b; \mesz, Rees, \& Wijers 1997; Panaitescu, \mesz,
\& Rees 1998).  The best motivated choices of $g$ are $g=0$ and
$g=2$, which correspond to a uniform density medium and the
density profile expected from a constant speed wind from the burst
progenitor expanding into a vacuum (or an ambient medium of much lower
density).  When the ambient density decreases with increasing distance
from the burster, the general result for spherical symmetry is a
faster decay of the afterglow flux (e.g. Panaitescu et al 1998)
(though the duration of the afterglow should increase correspondingly).
We therefore infer that a decreasing density profile will also
steepen the light curve decline in the beamed case.  This only 
exacerbates the disagreement between the observed slow afterglow decay
and the model predictions for beamed bursts, given the observed
spectral energy distribution.

It is worth asking how the exponential regime of burst remnant
evolution (section~\ref{exp_dyn_analytic}) will change if the ambient
density is nonuniform.  The exponential scale length $\rg
\propto \rho^{-1/3}$ in the uniform density case.  We therefore
conjecture that a solution similar to the following may be possible:
\begin{eqnarray}
\Gamma & \sim & \exp\left[ -\left( r \over \rg(\rho_1)
\times (\rho / \rho_1)^{-1/3} \right) \right] \\
\nonumber & \sim &
\exp\left[ -\left( r\over \rg(\rho_1) \times (r / r_1)^{g/3} \right) \right]
\end{eqnarray}
so that
\begin{equation}
\Gamma \propto \exp\left[ -\left( r \over \widehat{\rg} \right)^{(3-g)/3
} \right]
\end{equation}
where $r_1$ and $\rho_1$ are some fiducial radius and the
corresponding density, and where $\widehat{\rg} = \left( \rg(\rho_1)^3 \,
r_1^{-g} \right)^{1/(3-g)}$. 

\subsubsection{Observational concerns}
Throughout this discussion, we have tacitly assumed that the spherical
fireball does fit the GRB 970508 observations well, so that
difficulties with the various beaming models offer support to the
spherical model.  This is open to question.  In defense of the
spherical model, Reichart (1997) has studied the largest available
afterglow data set from a single optical observatory (that of Sokolov
et al 1997) and finds that the afterglow of GRB 970508 is well fitted
using a standard model with the addition of a modest amount of dust
extinction at high redshift.  By using a single data set, he minimizes
many of the possible systematic errors, such as inconsistent zero
points for absolute photometry.

On the other hand, if one examines the spectral slope from mixed data
sets over larger wavelength intervals (optical - near infrared) and a
larger time range, we find some worrying data points.  In particular,
the HST observations (Pian et al 1998) exhibit a spectral slope
$\alpha \approx 1.5 \pm 0.3$ based on quoted R ($0.7 \micron$) and H
($1.6 \micron$) band magnitudes
from the STIS and NICMOS instruments.  The observed slope in Sokolov et al's
data is $1.1$, which Reichart interprets as a reddened spectrum with
intrinsic slope $0.8$.  The HST data point is thus in mild conflict
with the Sokolov et al observation.  The significance of this conflict
is unclear, since the error on the HST data is dominated by
calibration uncertainties.

%

Likewise, if we compare the spectral slopes inferred from the Keck
K$_s$ band data (Morris et al 1997) and near contemporaneous optical
data (Djorgovski et al 1997; Keleman 1997), we find slopes of $0.24
\pm 0.12$ at $\tE = 4.35 \day$ and $0.40 \pm 0.10$ at $\tE = 7.35
\day$.  These are now substantially bluer than the value from Sokolov
et al.  The first of these may simply indicate that the K band flux
has not yet passed its peak and entered the power law decay phase;
scaling from the R band peak at $\tE\approx 1.9 \day$ gives a K band
peak at $\sim 4.1 \day$.  The second is harder to explain physically
but easier observationally, because the R band flux estimate is based
on an unfiltered observation that may have (for example) a substantial
color term.

The net effect of such outlying data points is illustrated by the
light curve fits in Rhoads 1999b.  These fits achieve $\chi^2$ per
degree of freedom around $3.6$ in fitting to a large compilation of R
band data (Garcia et al 1998).  It is not likely that any current
model can do better without discarding either predictive power or
outlying data points.

In summary, the spherical model does fit the GRB 970508 afterglow
model better than the beamed model developed in this paper for any
beaming angle $\zeta_m < 0.5$ radian.  There are
a few discrepancies in the measured spectral slopes.  If these are
real, they pose a challenge to standard fireball models.  However,
they could merely be indicative of calibration problems in
inhomogeneous data.  It is noteworthy that Sokolov et al (1997), who
have the largest multiband data set from a single telescope, find
no evidence for spectral slope evolution over the interval $2 \day \le
\tE \le 5 \day$.

\section{Conclusions} \label{summary}
We have shown that under a simple model of beamed gamma-ray bursts,
the dynamical evolution of the burst remnant changes at late times.
This change introduces a break in the light curve, which is
potentially observable.  The afterglows of GRB 970508, 970228, and
971214 showed no convincing evidence for such breaks, and their
combined spectral slopes and light curves are inconsistent with the
predictions of this beamed model.  This implies that beaming tests
based on blind searches for afterglows must be prepared to identify
transients whose properties differ appreciably from the properties of
these approximately spherical afterglows.  Some more recent afterglows
(GRB 990123; GRB 980519) better match beamed burst models, and may
provide more suitable templates for these searches.

Comparing our model with late time optical and radio observations, we
suggest that GRB 970508 was effectively unbeamed.  This implies 
energy requirements that are near the canonical isotropic values
for cosmological distances, and are not greatly 
mitigated by strong beaming.  No straightforward variation on our
beamed fireball model seems likely to simultaneously explain the
observed spectral slope and a pure power law light curve decaying at
the observed rate.  We conjecture that strongly beamed fireball models cannot
explain all observed gamma ray burst afterglows without substantially
altering at least one major ingredient of the models.  We therefore
obtain the first lower bound on GRB energy requirements that does not
involve assumptions about beaming: $E \ga 3 \times 10^{49} \erg$.
This limit will increase by an order of magnitude if the same material
that gives rise to the optical afterglow causes either the X-ray
afterglow or the gamma ray emission, and will rise further if the
burst's energy is not converted to radiation with perfect efficiency.


\acknowledgments
I wish to thank Ralph Wijers, Jonathan Katz, Tsvi Piran, Eli Waxman,
Daniel Reichart, Alexander Kopylov, David De Young, and Sangeeta
Malhotra for helpful communications.  I also wish to thank Infrared
Processing and Analysis Center for hospitality during the course of
this work.  Finally, I wish to thank all those who worked to achieve
accurate gamma ray burst position measurements and so opened the way
for afterglow studies.  This work was supported by a Kitt Peak
Postdoctoral Fellowship.  Kitt Peak National Observatory is part of
the National Optical Astronomy Observatories, operated by the
Association of Universities for Research in Astronomy.


\end{document}